\documentclass[aoas,preprint]{imsart}

\RequirePackage{amsthm,amsmath,amsfonts,amssymb,float,booktabs,pdflscape,bm}
\RequirePackage[authoryear,longnamesfirst]{natbib}
\RequirePackage[colorlinks,citecolor=blue,urlcolor=blue]{hyperref}
\usepackage[dvipsnames]{xcolor}
\RequirePackage{graphicx}% uncomment this for including figures

\startlocaldefs

\makeatletter
\newcommand*{\addFileDependency}[1]{ % argument
  \typeout{(#1)}
  \@addtofilelist{#1}
  \IfFileExists{#1}{}{\typeout{No file #1.}}
}
\makeatother

\endlocaldefs

\begin{document}

\begin{frontmatter}
\title{How Close and How Much?\\
Linking Health Outcomes to Built Environment Spatial Distributions}
\runtitle{Linking Health Outcomes to Built Environment Spatial Distributions}

\begin{aug}
%%%%%%%%%%%%%%%%%%%%%%%%%%%%%%%%%%%%%%%%%%%%%%
%%Only one address is permitted per author. %%
%%Only division, organization and e-mail is %%
%%included in the address.                  %%
%%Additional information can be included in %%
%%the Acknowledgments section if necessary. %%
%%%%%%%%%%%%%%%%%%%%%%%%%%%%%%%%%%%%%%%%%%%%%%
\author[A]{\fnms{Adam T.} \snm{Peterson}\ead[label=e1]{atpvyc@umich.edu}},
\author[B]{\fnms{Veronica J.} \snm{Berrocal}},
\author[C]{\fnms{Emma V.} \snm{Sanchez-Vaznaugh}}
\and
\author[D]{\fnms{Brisa N.} \snm{S\'anchez}}

%%%%%%%%%%%%%%%%%%%%%%%%%%%%%%%%%%%%%%%%%%%%%%
%% Addresses                                %%
%%%%%%%%%%%%%%%%%%%%%%%%%%%%%%%%%%%%%%%%%%%%%%
\address[A]{Department of Biostatistics, University of Michigan-Ann Arbor, \printead{e1}}
\address[B]{Department of Statistics, University of California-Irvine}
\address[C]{Department of Public Health, San Francisco State University}
\address[D]{Department of Biostatistics and Epidemiology, Drexel University}
\end{aug}

\begin{abstract}
Built environment features (BEFs) refer to aspects of the human constructed environment, which may in turn support or restrict health related behaviors and thus impact health. In this paper we are interested in understanding whether the spatial distribution and quantity of fast food restaurants (FFRs) influence the risk of obesity in schoolchildren. To achieve this goal, we propose a two-stage Bayesian hierarchical modeling framework. In the first stage, examining the position of FFRs relative to that of some reference locations - in our case, schools - we model the distances of FFRs from these reference locations as realizations of Inhomogenous Poisson processes (IPP). 
With the goal of identifying representative spatial patterns of exposure to FFRs, we model the intensity functions of the IPPs using a Bayesian non-parametric viewpoint and specifying a Nested Dirichlet Process prior. The second stage model relates exposure patterns to obesity, offering two different approaches to accommodate uncertainty in the exposure patterns estimated in the first stage: in the first approach the odds of obesity at the school level is regressed on cluster indicators, each representing a major pattern of exposure to FFRs. In the second, we employ Bayesian Kernel Machine regression to relate the odds of obesity to the multivariate vector reporting the degree of similarity of a given school to all other schools.
Our analysis on the influence of patterns of FFR occurrence on obesity among Californian schoolchildren has indicated that, in 2010, among schools that are consistently assigned to a cluster, there is a lower odds of obesity amongst 9th graders who attend schools with most distant FFR occurrences in a 1-mile radius as compared to others.
\end{abstract}

\begin{keyword}
\kwd{Food environment} 
\kwd{Child Obesity}
\kwd{Inhomogenous Poisson Process}
\kwd{Density Estimation}
\kwd{Nested Dirichlet Process}
\end{keyword}

\end{frontmatter}
%%%%%%%%%%%%%%%%%%%%%%%%%%%%%%%%%%%%%%%%%%%%%%
%% Please use \tableofcontents for articles %%
%% with 50 pages and more                   %%
%%%%%%%%%%%%%%%%%%%%%%%%%%%%%%%%%%%%%%%%%%%%%%
%\tableofcontents

%%%%%%%%%%%%%%%%%%%%%%%%%%%%%%%%%%%%%%%%%%%%%%
%%%% Main text entry area:

\section{Introduction}
\label{sec:Intro}

The dramatic increase in child obesity is one of the most pressing public health issues of the 21st century \citep{world2012population}. The potential causes of lack of energy balance that result in child obesity have been widely studied, and the need for population-level interventions, beyond individual-level treatments has been strongly emphasized by the research community and policy makers alike \citep{IOM}. Place-based interventions are one realm of population level approaches that seek to modify neighborhood environments in ways that can support residents' health promoting behaviors. Within this type of approach, changes to the distribution of health-supportive (or detrimental) amenities within neighborhood environments have emerged as a possibility, given that the built environment -- the human made space in which humans live, work and recreate on a day-to-day basis -- constrains everyday health-relevant choices \citep{roof2008public}. 

In particular, the potential contribution of the food environment near schools (e.g., fast food restaurant availability) to child obesity has been studied extensively  \citep{currie2010effect,davis2009proximity,sanchez2012differential,baek2016distributed}, given that children spend large proportions of their waking hours and consume a large proportion of their food within and near the school environment.  While the body of evidence supports these connections broadly, different approaches to conceptualize exposure make it challenging to more fully understand the health effects of environmental exposures, as well as identify where interventions may be especially needed. In order to assist policy makers with these challenges, methods need to be developed that both (i) identify different spatial patterns of exposure and (ii) link these patterns to health outcomes quantitatively. Exposure patterns, compared to continuous exposure measures, may make it more straightforward to identify places in higher need of interventions.
\par 
Previous work in environmental epidemiology has approached these problems by first clustering some measure of built environment features (BEFs), e.g. the number of fast food restaurants (FFR) within a mile, and then incorporating cluster assignments as a categorical predictor into a second stage regression model. For example, \cite{wall2012patterns} used a spatial latent class analysis (LCA) to cluster multivariate measures of the built environment, including the density of food outlets within 1 mile of the subjects residential location, and subsequently estimated the association between cluster membership and adolescent obesity. Like \cite{wall2012patterns}'s analysis, it is common to use a simple count of BEFs within some pre-specified buffer (e.g. 1 mile) as an exposure measure (e.g., \cite{an2012school,howard2011proximity}). 
%However, this assumes that the effect of each BEF is identical across the buffer, an assumption that is not empirically supported \cite{baek2016distributed,fotheringham1991modifiable} \textcolor{orange}{not completely sure about this statement. I think we need to be careful about spatially-varying covariates and spatially-varying coefficients, which it seems to me this sentence is referring to}. 
However, clusters identified with these traditional exposure summaries ignore the spatial distribution of amenities within the buffer. This spatial distribution is relevant from a mechanistic perspective because BEFs closer to schools are easier to access, as well as policy relevant since the distribution could inform built environment interventions such as zoning laws to curtail exposure.  Finally, plugging in an estimate of cluster membership as a predictor in a health outcome model does not account for the uncertainty in the estimated cluster assignment label, leading to potentially incorrect inference of the associated health effect.

Motivated by the need to better understand the association between exposures to FFRs near schools and child obesity, this paper has two complementary goals. First, we aim to develop a clustering procedure that provides interpretable groupings of BEF exposure while taking into account the spatial distribution of BEFs.  For this goal, we work with the geographical coordinates of BEFs and schools, modeling the set of distances of each school to its nearby BEFs as a realization of a 1-dimensional point pattern process with a school-specific intensity function. Clusters of schools are formed by clustering the intensity functions of these point patterns using a Nested Dirichlet Process (NDP). Working with point-level data and using the actual distances between schools and BEFs, rather than aggregating the data at the areal-unit level, allows us to maintain the level of granularity needed to investigate the effect of the spatial distribution of BEFs around schools on children's obesity. In particular, our approach to deriving the schools' cluster assignments is based on the distribution of distances of schools and their surrounding FFRs, but not the quantity of FFRs. This clustering approach enables us to separate the contribution to obesity associated with the number of FFRs near schools from the association of obesity with the relative proximity of FFRs to the school, thereby providing new insights compared to prior work. Second, we show two ways to use the output from the NDP clustering model to address cluster assignment uncertainty when using clusters as predictors in a regression that evaluates the association between FFR exposure near schools and obesity risk of students in those schools.   
 \par 

As a statistical genre, clustering methods vary widely, from the model-based finite mixture models (FMM) \citep{diebolt1994estimation} and previously mentioned LCA \citep{wall2009spatial}, to the more algorithmic k-means style methods \citep{hartigan1975clustering,friedman2001elements}. Each of these have varying strengths and weaknesses according to the problem at hand. Notably, FMMs, K-means and LCA share the important assumption of pre-specifying the number of clusters that should be found in the data. LCA and K-means also make parametric assumptions about the relevant distribution or metric, respectively, that should define the clusters. In our pursuit of examining the contribution of both the spatial distribution and quantity of FFRs around schools on obesity without strong parametric assumptions or pre-specification of the number of clusters we employ a NDP approach. We use the NDP to flexibly cluster schools according to the  spatial distribution of BEFs around them, which we assume is regulated for each school by the intensity function of an Inhomogenous Poisson Process (IPP). The NDP allows us to estimate these functions without enforcing any strong parametric constraints on the shape of the intensity function or the number of clusters. Akin to \cite{xiao2015modeling}, our model for the intensity function of the IPP factorizes the intensity function into the product of a normalized intensity function modeled non-parametrically, and a total intensity. Nevertheless, our model is different from that of \citet{xiao2015modeling} in multiple ways. First, whereas we are dealing with an IPP over space with the goal of identifying common patterns in the spatial distribution of FFRs around schools, while the latter work with a marked IPP over time in order to examine the inter- and intra-annual variation of hurricanes frequency. Additionally, while \citet{xiao2015modeling} invoke a dependent Dirichlet process (DDP) \citep{maceachern1999,maceachern2000} to capture the temporal dependencies across years in hurricanes' frequencies when modeling the intensity function, our model employs a Nested Dirichlet Process to identify clusters of schools with similar spatial distributions of FFRs near them.  

Additionally, in accordance with our conceptual objective (i), the NDP provides cluster assignment labels  which can be processed and used in a second-stage regression analysis to estimate associations between the BEF's spatial distribution and a health outcome of interest. Second-stage models raise the  need to accommodate uncertainty in estimated exposures (in this case cluster assignment), a need that has been the topic of several papers \citep{chiang2017hierarchical,graziani2015bayesian,wall2012patterns,wade2018bayesian}. Here, we explore two approaches to using the output of our clustering model in a second-stage analysis as a way to deal with the challenges of making cluster assignments; namely the NDP yields a varying number of cluster assignments in the posterior samples. One approach relies on using a conservative ``consensus of cluster assignments'' as determined from cluster-specific uncertainty bounds \citep{wade2018bayesian}. The second approach avoids the use of a single cluster assignment by using each school's vector of co-clustering probabilities with other schools as a measure of multivariate exposures, inserting the vector as a predictor  into a health outcome model through a Bayesian kernel machine (BKMR) regression approach \citep{bobb2015bayesian,valeri2017joint,coker2018association,wang2018associations}. This latter approach is an innovation in terms of expanding the applications of BKMR, as well as a way to utilize a clustering model's output to address classification uncertainty. 
\par 
The layout of the paper is as follows: Section 2 discusses the data sources used in our analysis of child obesity in relation to FFR occurrence near their schools, namely data on children's obesity status and school characteristics obtained from the California Department of Education, and food outlet locations from the National Establishment Time Series (NETS) database. The section includes discussion of some nuances involved in handling this and similar data for our proposed statistical methodology, as well as some preliminary data analysis. Section 3 describes the modeling approach that we propose for clustering schools with respect to the spatial distribution of FFRs around the school and the two modeling frameworks for the second stage health analysis.  Section 4 contains the results from fitting our models to the California data. We finish with a discussion of the contribution our work makes to the built environment literature, limitations of our approach and possible methodological extensions.

\section{Data on child obesity and food environment near schools in California}
\label{sec:Data}

\subsection{Data sources and study sample}
\label{sec:Data_sources}
Each spring, public schools in the State of California collect data on the fitness status of pupils in 5th, 7th and 9th grade, as part of a state mandate, using the Fitnessgram battery. The Cooper Institute's sex-, age- and height-specific standards for body weight are used to classify each child as "meeting the standard", "needs improvement", or "needs improvement, high risk", which correlate to normal, overweight, and obese classifications. We use the last two of these as "not meeting the standard", and use the term obesity henceforth when referring to this outcome. Fitnessgram data are available through the California Department of Education (CDE) website (https://www.cde.ca.gov/ds/). In our analysis, we use data collected during academic year 2009-2010 on 9th graders only, since high school youth are more likely to be exposed to the food environment surrounding their school (e.g., students may leave the campus for lunch). 

Data on school-level characteristics are also available from the CDE website (see Table \ref{tab:CA_descriptive}), and, importantly, so is the geocode of the school. School geocodes were used for two purposes. First, the geocodes were used to link schools to census tract level covariates. Second, the school geocodes were used to calculate the distances between the school and the geocoded location of each FFR in California. FFRs were identified from the National Establishment Time Series (NETS) database \citep{walls2013national}, using a published algorithm that classifies specific food establishments as FFRs \citep{auchincloss2012improving}. Only distances shorter than one mile were kept for this analysis. This distance was chosen on the basis of previous work that estimated that the distance at which FFRs cease to have an effect on childhood obesity is approximately one mile \citep{baek2016distributed}.  Finally, we calculated the distance between all schools, to derive a data set of schools so that there are never two schools within one-mile of one another, in order to satisfy independence assumptions used in the analysis. 
\par
 
\par 
\subsection{Preliminary analysis}
\label{sec:prelim}
The dataset is comprised of 420,085 children who attended 1,193 high-schools. Across all high schools, 40\% of the 9th graders were observed to be obese. Of the high-schools, 767 had at least one FFR within one mile and 426 had zero.  Although the second stage analysis for the health outcome includes \emph{all} schools in the dataset, regardless of whether they have FFRs or not within a mile, the first stage analysis that derives clusters of school with similar spatial distribution of surrounding FFRs excludes the 426 schools that do not have any FFRs within one mile of their location. \\

Descriptive statistics of the schools are presented in Table \ref{tab:CA_descriptive}, for the entire dataset and for the two subsets of schools without FFRs or with at least one FFR within a mile.  As the Table shows, aside from having at least one FFR, schools included in the first stage analysis are generally more likely to be located in urban areas (46$\%$) compared to schools not included (27$\%$).   
Schools included in the first stage analysis varied both in terms of the number of nearby FFRs and in terms of their spatial distribution, both important aspects of BEF exposure. Among these schools, 45\% had 1 to 4 FFRs within a 1-mile buffer while the rest of the schools had at least 5; additionally the median (Q1-Q3) distance to the first FFR was 0.4 (0.3-0.6) miles. 

\begin{figure}[H]
    \centering
    \caption{Panel A: Distribution of distances from the school to nearby FFRs for two schools with 10 fast food restaurants (FFRs) within a 1 mile radius. Panel B: Distribution of distances from the school to nearby FFRs for two schools that have the same distance to the  closest FFR. Panel C: distribution of distances to FFRs for a sample of 100 schools.  For each school the plot shows the range of distances between the 2.5th and the 97.5th percentile. Schools are sorted by median distance to FFR. Darker dashed and dotted lines represent the four schools depicted in panels A and B of this figure. }
    \includegraphics[width = .9\textwidth]{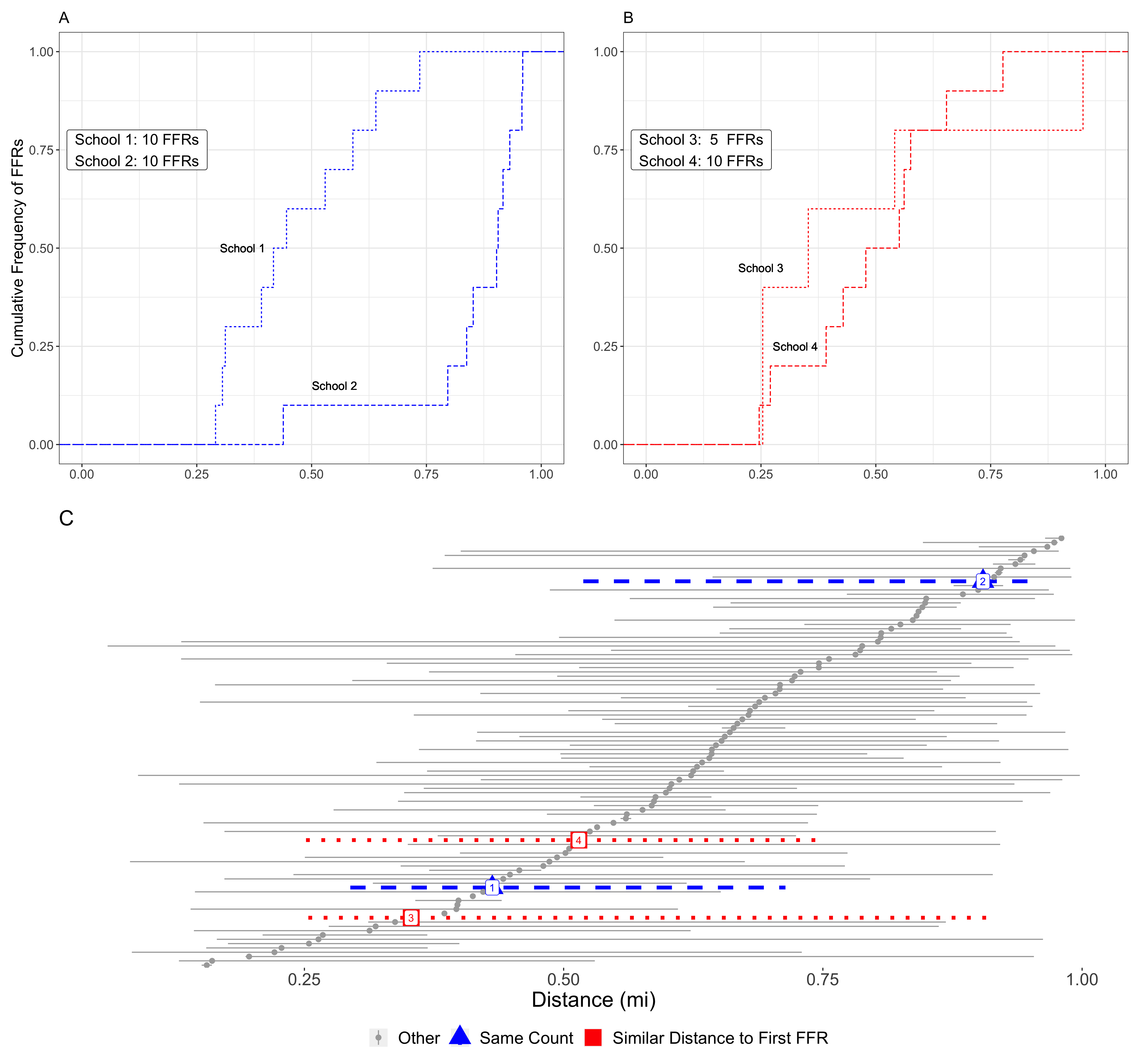}
    \label{fig:composite_distance}
\end{figure}

A richer understanding of schoolchildren's exposure to FFRs can be gleaned by examining the empirical cumulative distribution function (ECDF) of the distances between each school and its neighboring FFRs.  The ECDFs for four schools are shown in the top panels of Figure \ref{fig:composite_distance}. These two panels illustrate how traditional measures of built environment exposure, using simple counts or distance to the closest FFR, may fail to incorporate meaningful aspects of spatial exposure.

\begin{table}[H]
    \centering
\begin{tabular}{llccc}
\toprule
 &  & \multicolumn{2}{c}{Subset of schools with} & \multicolumn{1}{c}{All schools} \\
 &  &  $\geq 1$ FFR nearby & no FFRs nearby & \\ \toprule
\multicolumn{1}{l}{\underline{Children}}\\ 
Number       &  & 298,903 & 121,182 & 420,085 \\
$\%$ Obese   &  &  40$\%$  & 40$\%$  & 40$\%$ \\
\midrule
\multicolumn{1}{l}{\underline{Schools}}\\ 
Number & & 767$^{\star}$ & 426 & 1,193 \\
Urbanicity & Rural &  $10$ & $39$ & $21$ \\
 & Sub-Urban &  $44$ & $34$ & $40$ \\
 & Urban &  $46$ & $27$ & $39$ \\
Majority Race & African American &  $\phantom{0}1$ & $\phantom{0}2$ & $\phantom{0}1$ \\
 & Asian &  $\phantom{0}4$ & $\phantom{0}2$ & $\phantom{0}4$ \\
 & Hispanic &  $29$ & $27$ & $28$ \\
 & No Majority &  $14$ & $\phantom{0}9$ & $12$ \\
 & White &  $53$ & $59$ & $55$ \\
Median income (\$1,000 USD)  & Median (Q1-Q3)  & 59.1 (43.4-78.6) & 52.3 (40.2-74.4) & 56.7 (42.5-77.1) \\
of households  in school's census tract & IQR  & $35.2$ & $34.1$ & $34.6$ \\
Proportion of residents  & Median (Q1-Q3)  & 23.7 (12.6-38.6) & 19.6 (11.2-32.6) & 22.2 (12.1-36.2) \\
with $\geq$ 16 years of education & IQR  & $26.0$ & $21.4$ & $24.1$ \\
FFR Quantity &0 & $\phantom{0}0$ & $100$ & $36$ \\
 & [1,4] &  $45$ & $\phantom{0}0$ & $29$ \\
 & $\geq$5 & $55$ & $\phantom{0}0$ & $36$ \\
 Closest FFR (Miles) & Median (Q1-Q3)  & 0.4 (0.3-0.6) & - & 0.4 (0.3-0.6) \\
 & IQR  & $\phantom{0}0.4$ & - & $\phantom{0}0.4$ \\
\bottomrule
\end{tabular}
    \caption{Descriptive statistics for children and schools in the analytic dataset. Summary statistics for categorical variables are the percentage of data in the column. IQR = Inter-quartile range.
    $^\star$17 schools have missing data on obesity.}
    \label{tab:CA_descriptive}
\end{table}

Specifically, Figure \ref{fig:composite_distance}A illustrates how schools that may have a similar number of FFRs within a given distance may be characterized by a dramatically different spatial distributions of the surrounding FFRs. Likewise, Figure \ref{fig:composite_distance}B illustrates that while certain schools may have similar distribution of distances to FFRs, the total number of nearby FFRs might be completely different. Characterizing exposure to FFRs on the basis of these traditional measures or clustering schools based only on how similar these two exposure metrics are across schools could miss meaningfully important aspects of spatial exposure, and consequently not fully capture the effect of exposure to unhealthy food on health. Figure \ref{fig:composite_distance}C illustrates the distribution of school-FFR distances for 100 randomly selected high schools, including the aforementioned four illustrative schools, further demonstrating the wide variability in exposure to FFRs in our dataset.  In order to accommodate the limitations in capturing exposure to FFRs discussed above, our modeling approach considers \emph{all} the distances from each school to its nearby FFRs, modeling this set of distances as a realization of a 1-dimensional point pattern process whose school-specific intensity function is estimated non-parametrically.  

\section{Model}
\label{sec:Model}

As mentioned in Section~\ref{sec:Intro}, our analysis of the effect of exposure to FFRs on obesity is based on a two-stage approach: (i) a first stage model that characterizes the main patterns of school-level exposure to FFR, by deriving characteristic profiles of the spatial distribution of FFRs near schools; (ii) a second stage model that uses the output from stage 1 in a regression model to examine the association between patterns of exposure with child obesity. In the following sections we provide details of each of these modeling strategies, including specific aspects of our models' implementations and estimation.

\subsection{Clustering model}
\label{sec:mod_cluster}
To characterize the food environment near schools, our clustering approach focuses on the point processes describing the relative locations of the FFRs in the immediate vicinity of the schools, rather than the global 2-dimensional point process representing the location of FFRs across the entire state of California.  Specifically,  let $r_{ij}$ be the distance between the $i$th school $(i=1,...,N)$ and the $j$th nearby FFR $(j=1,...,n_i)$: each $r_{ij}$ belongs to the interval $(0,R) \subset \mathbb{R}$, with maximum distance $R$ chosen on substantive grounds. Since the schools in the sample are relatively far from each other (by at least $R$), the distribution of distances for one school does not inform on the distribution of distances for another. Thus,  for each school $i$, we model the random subset $\mathcal{D}_i = \{r_{ij} ;j =1,...,n_i\}$ as a realization from a one-dimensional Inhomogeneous Poisson Process (IPP) with intensity function $\lambda_i(r), r\in (0,R)$. We further decompose the intensity function $\lambda_i(r)$ as $\lambda_i(r)=\gamma_i f_i(r)$ with $\gamma_i$ representing the expected number of FFRs within radius $R$ from school $i$ and $f_i(r)$ denoting a normalized density. Thus, the $i$th school's contribution to the likelihood is: 
\begin{displaymath}
p\left( \mathcal{D}_i | \gamma_i,f_i(r)  \right) \propto  \gamma_i^{n_i}\exp\{-\gamma_i\} \prod_{j=1}^{n_i} f_i(r_{ij}). \label{eqn:mod_llhood} \tag{1}    
\end{displaymath}
Assuming independence between schools, the full likelihood is obtained by taking the product over the $N$ schools' likelihood contributions. In (\ref{eqn:mod_llhood}) we note that the  likelihood is separated into a component that handles the number, $n_i$, of FFRs for each school $i$, and a component that, given $n_i$ FFRs surrounding school $i$, evaluates the density at each of the $n_i$ distances. For our purposes, $\gamma_i,i=1,...,N$ are considered nuisance parameters that do not affect the estimation or interpretation of the $f_i(r)$ beyond what has been previously discussed. In the health outcome model we use the observed $n_i$'s directly as a predictor, instead of their expected values $\gamma_i$'s, in accordance with our aim to differentiate between the separate effects on childhood obesity of the observed FFR quantity and the FFRs' spatial distribution.

\par Our goal is to simultaneously model and cluster the FFR spatial density functions $f_i(r)$, $i=1,\ldots,N$, in a non-parametric fashion.
The non-parametric estimation of a single $f_i(r)$ could be accomplished by using a Dirichlet Process (DP) mixture model \citep{gelman2013bayesian}. However, in order to fulfill our goal of non-parametrically estimating and clustering the $f_i(r)$'s themselves, we use a NDP modeling approach.  The NDP enables us to achieve both of these objectives through the use of two DP's simultaneously. Specifically, we express each $f_i(r)$ as: 
\begin{align*}
    f_i(r) &= \int \mathcal{K}(r|{\bm{\theta} })dG_i(\bm{\theta}) \label{eqn:mod_ndp1} \tag{2}\\
    G_i &\stackrel{iid}{\sim} Q\\
    Q &\sim DP(\alpha,DP(\rho,G_0)),
\end{align*}
where, $\mathcal{K}(r|\bm{\theta})$ is a mixing kernel with  parameter vector $\bm{\theta}$, and the distribution $G_i$ is drawn from the random distribution $Q$ on which we place a NDP prior. In (2), $DP(\rho,G_0)$ denotes a DP with concentration parameter, $\rho>0$, and parametric base measure, $G_0$. The base measure $G_0$ is the distribution around which the DP is centered and the concentration parameter, $\rho$, reflects the variability around that base measure.
\par It is through the $G_i$'s that the $f_i(r)$'s are clustered as can be seen from the stick breaking construction representation of the NDP: $Q = \sum_{k=1}^{\infty} \pi_k^*\delta_{G_k^{*}}(\cdot)$. In this representation, $\pi_k^*$ represents the probability that a school is assigned to the $k$-th mixing measure, $G^{*}_k$, $\delta_{\cdot}(\cdot)$ is the Dirac delta function and $G_k^*= \sum_{l=1}^{\infty} w_{lk}^{*} \delta_{\bm{\theta}^{*}_{lk}}(\cdot)$ is itself composed of weights, $w_{lk}^*$ and associated atoms $\bm{\theta}_{lk}^*$. This hierarchy of distributions, weights and atoms provides a framework that flexibly identifies clusters of schools, and also flexibly estimates the intensity function representing the spatial distribution of FFRs surrounding schools for each cluster. 

Combined altogether, the hierarchical formulation of our model is:
\begin{align*}
\{r_{ij} ;j =1,...,n_i\} &\stackrel{ind}{\sim} IPP(\lambda_i(r)), \quad i = 1,\ldots,N \label{eqn:mod_ndp2}\tag{3}\\
%\{r_{ij}\}_{j=1}^{n_i} &\sim IPP(\lambda_i(r)), \quad i = 1,...,N\\
\lambda_i(r) & = \gamma_i f_i(r)  \\
    f_i(r) &= \int \mathcal{K}(r|\bm{\theta}) dG_i(\bm{\theta})\\
    G_i & \stackrel{iid}{\sim} Q\\
    Q & \sim DP(\alpha, DP(\rho,G_0)),
\end{align*}
\noindent and, as previously noted, the $\gamma_i$ are nuisance parameters that do not influence the estimation of the intensity functions, which are of primary interest.

\par
In our analysis of FFR exposure around California public high schools we transform the school-FFR distances from $(0,R) \to \mathbb{R}$ using a probit function to create the modified distances $r'_{ij}=\Phi^{-1}(r_{ij}/R)$. We make this transformation in order to use a normal mixing kernel and corresponding normal-inverse-chi square base measure, $G_0 = N(0,\sigma)\times\text{Inv}-\chi^2(1,1)$, in order to facilitate computation. Similarly, we place informative Gamma priors, on the concentration parameters, $\alpha,\rho \sim \text{Gamma}(10,10)$, to encode our \textit{a priori} belief that there should be a small number of clusters, in line with similar work \citep{ishwaran2001gibbs,gelman2013bayesian,rodriguez2008nested}.

\subsection{Health Outcomes Model}
\label{sec:mod_outcomes}
In the second stage of the analysis, we examine whether spatial distributions of FFRs around a school are associated with obesity of children in the school. For this, we use results from the clustering model (3) as input in to a regression model where child obesity is the outcome.

 \par The simplest way to do this would to be to assign a cluster label to each school and use it as a covariate in the health outcome regression. However, proceeding in this fashion would not account for the uncertainty in the cluster assignment and would not exploit all the information in the posterior samples that are generated while fitting the clustering model of Section \ref{sec:mod_cluster}. To address these issues, we propose two alternative approaches.
 
 \subsubsection{Consensus generalized linear model (CGLM)}
  The first approach \emph{controls} uncertainty in the cluster labels by using in the health outcome model only the schools for which the cluster label is known with greater relative certainty, as follows.  First, we derive cluster assignment labels for each school using the posterior samples and a loss function in a decision theoretic framework \citep{wade2018bayesian}. Specifically, we use the variation of information (VI) loss function to determine the optimal cluster configuration, which simultaneously identifies both  the number of clusters and cluster labels for the schools. This approach finds the posterior sample that produces the minimal loss, and uses the number of clusters and cluster assignments in that sample to assign labels to schools-- thus deriving, essentially, a ``point estimate'' for the discrete/categorical cluster assignment. We refer to this point estimate as the `mode' cluster label. \\
 
 Second, we identify schools with low uncertainty in the cluster label. In addition to the mode cluster configuration, the method also produces 95\% uncertainty bounds for both the number of clusters and for the cluster labels for each school, yielding three additional cluster configurations (for a total of four including the mode). Compared to typical ``upper'' and ``lower'' bounds, the method provides three bounds according to the combination of loss metric value and the number of clusters, the latter of which can vary from one configuration to another.   Since, as noted earlier, employing the cluster labels from the single point estimate as a predictor in the health outcome model would ignore the uncertainty associated with the cluster assignment label, our goal here is to take into account all four cluster configurations to control or reduce the uncertainty. One possibility is to use each of the four assignments in separate health outcome regression models, and subsequently fuse together their results. However, fusing those four models may entail fusing models with potentially a different number of clusters. Thus, rather than using each of the four alternative cluster labels in the health outcome models, we restrict the health outcome analysis to the set of schools that are assigned to the same cluster across the four different cluster assignments. Note that this is only possible when clusters are well identified and posterior samples do not exhibit label-switching across iterations -- as is our case -- or a post-processing step that adjusts for label switching has been run \citep{gelman2013bayesian,rodriguez2014label,stephens2000dealing,panagiotis}. These conditions ensure that cluster labels are consistent across configurations and, consequently, taking the intersection has a consistent meaning. \\  
 
In summary, the CGLM approach addresses the uncertainty in the cluster label assignment by taking the intersection of the four cluster labels to arrive at a more conservative (less uncertain) estimate of the schools' cluster assignment. This reduction of uncertainty in cluster assignment comes at the cost of sample size, as schools will be included or excluded from the health regression model according to whether they fall within said intersection or not. Despite this loss of sample size, the key advantages of this approach are that this enables a straightforward analysis, as the intersection of the four cluster configurations yields a single cluster assignment for each school that can be used as a categorical covariate in the health outcome model, and the cluster assignment is more precise than the single ``point estimate'' label in the entire sample.\\ 
 
To define the CGLM outcome model and enable us to distinguish the association between the quantity of FFRs and obesity from that of the FFRs' spatial distribution, we bring back into consideration the schools that had zero FFRs within 1 mile. Let $C_{i,k} = I(\text{ith school belongs to cluster k})$, $k=1,...,K$, denote the cluster to which the $i$-th school is assigned ($K=6$ in our analysis). In addition, define $Q_{i,m}, m=0,...,M$ a set of indicator variables that treat the number of FFRs as a categorical variable. In particular $Q_{i,0}=I(n_i=0)$, with the other categories being $ n_i=2;n_i=3; n_i=4; n_i \in \{5, 6, 7\},$ and $n_i>7$.  This categorical representation is used given the distribution of FFRs and the lack of linearity in the association between FFR quantity and the odds of obesity. We note that the cluster indicators are only available for schools with $n_i>0$, and that the schools included in this model are those with  $n_i=0$ or $n_i>0$ that are determined by the consensus approach discussed above to have a cluster label with relatively higher certainty. This set of schools is denoted as $\mathbb{D}_{Consensus}$. \\

The CGLM model linking obesity in schoolchildren to exposure to FFRs and other school neighborhood characteristics is thus: 
 \begin{align*}
\text{logit}(\text{p}_{i'}) = \left\{ \sum_{m=0}^{M}Q_{i',m}\zeta_{m}\right\} + I(n_{i'}>0)\left\{ \sum_{k=1}^{K} C_{i',k}\xi_k \right\} + \bm{Z}_{i'}^{T} \bm{\beta} \qquad i' \in \mathbb{D}_{consensus} \label{eqn:outcomes_cglm} \tag{4}
 \end{align*} 
 
 In (\ref{eqn:outcomes_cglm}), p$_{i'}$ denotes the proportion of obese 9th grade students at the $i'$th high school, $\zeta_{m}$ and $\xi_k$ are quantity- and cluster-specific coefficients, respectively, and  $\bm{Z}_{i'}$ is a vector of school characteristics without an intercept term. Specifically $\bm{Z}_{i'}$ includes the categorical variable indicating the racial majority of the children enrolled in the school; an indicator denoting whether school $i'$ is a charter school; the school $i'$'s census tract median household income centered at the overall state median income and scaled by 33,000; the proportion of adults who have $\geq 16$ years of education within the school's census tract centered by the state census tract average; and the urbanicity level near the school, classified as rural, suburban or urban. Given the parametrization of the covariates, the reference category when $Z_i=0$ is a suburban high school, with a majority white student population, with the average percent of college educated adults and median census tract household income.

\subsubsection{Bayesian Kernel Machine Regression (BMKR)} In the second approach, we move away from using a categorical predictor in the health outcome model, and instead use BKMR, as explained below, to input the probabilities that any pair of schools belong to the same cluster into the health outcome model. This has several advantages compared to approaches that use a single set of cluster labels for schools (e.g., the mode) in the heath outcome model. When a categorical predictor,  denoting discrete groups of schools' FFR spatial profiles, is used in the health outcome model, we estimate the effect of each of these spatial profiles by borrowing information only from the schools within the same cluster. Hence, when we estimate health effects in this fashion,  we are only borrowing information from a limited number of schools, those that belong to the same disjoint subset of schools, i.e., the clusters. Moreover, as described above in the CGLM, additional steps are needed to account for cluster assignment uncertainty. In contrast to the CGLM specifically, which controls uncertainty by restricting the sample, the BKMR could utilize all the schools in the NDP.
 
 The BKMR approach proceeds as follows. First, to each school $i$, $i=1,\ldots,N$, in the clustering sample, we associate the $N$-dimensional row-vector $\mathbf{P}_i$, $i$-th row of the co-clustering probability matrix $\mathbf{P}$. This matrix is constructed by averaging, for each $i$, the indicators $I(\text{school}~i~\text{and}~j~\text{are in the same cluster})$, for $j\neq i$ across the posterior samples, with the $i$-th element of $\mathbf{P}_i$ set equal to 1 by convention. By using the $N$-dimensional vector $\mathbf{P}_i$ as a predictor in the health outcome model, we allow all schools, including those that  would not be assigned to the same cluster as school $i$, to contribute to the estimation of the effect of the spatial distribution of FFRs surrounding the school on its schoolchildren's odds of obesity.  In other words, in this approach we remove the hard boundaries that separate schools into disjoint subsets, and enable sharing of information about schools' FFR-spatial profiles across all schools, albeit the contribution of schools with more (compared to less) similar profiles is given higher weight. 

Second, rather than linking the log-odds of obesity at school $i$ to the $N$-dimensional row-vector $\mathbf{P}_i$, directly, in the logistic regression for obesity, we include as a \emph{predictor} the scalar $h(\mathbf{P}_i)$ for school $i$. Since schools that have large probabilities of being co-clustered are likely to have similar $N$-dimensional row-vectors, while schools that are not very likely to be co-clustered are associated with more dissimilar $N$-dimensional row vectors, we model the $N$ random terms  $h(\mathbf{P}_i)$, $i=1,\ldots,N$ jointly. Specifically, we model $\left( h(\mathbf{P}_1), \ldots,h(\mathbf{P}_N) \right)$ as a finite-dimensional realization of a Gaussian process with mean $\bm{0}$ and Gaussian covariance function $\kappa(\cdot, \cdot; \phi, \sigma^2)$.  The covariance function $\kappa(\cdot, \cdot; \phi, \sigma^2)$ is evaluated using as distance between any two $N$-dimensional row-vectors $\mathbf{P}_i$ and $\mathbf{P}_j$ the Euclidean distance, whereas the two parameters $\phi$ and $\sigma^2$ encode, respectively, the range of the correlation, e.g. the distance at which the correlation between any two $N$-dimensional random vectors is essentially negligible, and the marginal variance of each $h(\mathbf{P}_i)$. This approach borrows from the environmental epidemiological literature where researchers are often confronted with the issue of having multiple, potentially correlated, high-dimensional exposures.\\

As with the CGLM, we incorporate into the outcome model all observations with 0 FFRs within the mile, and use the indicators for quantity of FFRs nearby, $Q_{i,m}$, to distinguish the effect on obesity associated with the quantity of FFRs from their spatial distribution. Altogether, the second health outcome model is:
\begin{align*}
    \text{logit}(\text{p}_{i'}) &= 
    Q_{i',0}\zeta_{0}+ I(n_{i'}>0)\left\{\tilde{\alpha}  + h(\bm{P}_{i}) +  \sum_{m=0}^{M}Q_{i',m}\zeta_{m} \right\} + \bm{Z}_{i'}^{T} \bm{\beta}  \quad i' \in \mathbb{D}_{Full}  \label{eqn:outcomes_bkmr} \tag{5}\\
    h(\cdot) &\sim \mathcal{GP}(\bm{0},\kappa(\cdot,\cdot|\phi,\sigma^2)). 
\end{align*}
 In (\ref{eqn:outcomes_bkmr}), $\mathbb{D}_{Full}$ is the index set for schools with zero FFRs in addition to the full set of $N$ schools used in the first stage model. Additionally, $\tilde{\alpha}$ denotes the intercept for schools with at least one FFR, and $h(\bm{P}_{i'})$, $i' \in \mathbb{D}_{BKMR}$, indicates a school-specific random intercept. The other components of the model in (\ref{eqn:outcomes_bkmr}) - $\bm{\beta},\bm{Z}_{i'}$ - have the same definition and interpretation as before.
 
 \par For comparative purposes, we fit the BKMR to both datasets, $\mathbb{D}_{Consensus}$ and $\mathbb{D}_{Full}$. Similarly, we also fit a logistic regression to schools in the $\mathbb{D}_{Full}$ set using the same parametrization as in (\ref{eqn:outcomes_cglm}). In this latter case, the mode cluster assignment was used to determine cluster specific indicators. This model is hereafter referred to as the Mode GLM (MGLM). In all models, $\bm{\beta}$, $\zeta_m$,$m=1,...,M$ and $\xi_k$, $k=1,...K$ are given flat improper priors, while in the BKMR, $\phi$ and $\sigma^2$ are each given informative folded Normal$(1,3)$ priors to accommodate known identifiability issues \citep{zhang2004inconsistent}. These informative priors were chosen after initial runs with uniform priors on larger intervals of $\mathbb{R}^{+}$ for both parameters showed that posterior samples were contained in the $(0,1)$ interval.

\subsection{Estimation}
\label{sec:Estimation}

As both the clustering model and the health outcome models that we have proposed in Section \ref{sec:mod_outcomes} are specified within a Bayesian framework, inference on all model parameters are obtained through the posterior distribution, which we approximate using a Markov chain Monte Carlo (MCMC) algorithm. For our NDP clustering model we use the blocked Gibbs sampler as described in \citet{rodriguez2008nested}, truncating the summations for the inner and outer DPs using $L=30$ and $K=35$, respectively - a choice based on logic similar to that discussed in \cite{rodriguez2008nested}. This model fitting routine is implemented within our  \texttt{bendr} \citep{bendr} \texttt{R} package.  The health outcome models are fit using the Hamiltonian Monte Carlo variant sampler implemented in \texttt{stan} \citep{carpenter2016stan} via \texttt{rstan} (BKMR) \citep{rstan} and \texttt{rstanarm} (CGLM) \citep{rstanarm}. All model fitting was performed within \texttt{R} (v3.6.1) \citep{Rlanguage} on a Linux Centos 7 operating system with 2x3.0 GHz Intel Xeon Gold 6154 processors.
\par For the NDP model, 250,000 samples were drawn from the posterior distribution, with 240,000 iterations discarded for burn-in and the last 10,000 iterations thinned by 3 to reduce auto-correlation for a total of 3,333 posterior samples used for inference. The length of the burn-in period and thinning were determined by inspecting trace plots for various model parameters and by computing Raftery's diagnostic statistic \citep{raftery1995number}. 
For all health outcomes models, we ran 4 independent chains, using different initial values, each ran for 2000 iterations. For each chain, we kept 1000 samples after burn-in, for a total of 4000 posterior samples that we employed for posterior inference. Convergence was assessed via split $\hat{R}$ \citep{gelman2013bayesian} and visual inspection of trace plots. 
\par In fitting the proposed health outcome models, we exclude 17 schools previously included the clustering because they are missing outcome information. As discussed in Section \ref{sec:mod_outcomes}, the outcome models include an additional 446 high schools with no FFRs within the 1 mile radius, which were not considered in the clustering model. Cluster labels that are included as predictors in the CGLM and MGLM were derived using the \texttt{mcclust.ext2} package in \texttt{R} \citep{mcclust}. The same \texttt{R} package was employed to estimate the posterior assignment credible bounds with VI loss function as detailed in \cite{wade2018bayesian}. We take the intersection over the horizontal, upper and lower bounds as described in Section \ref{sec:mod_outcomes} to arrive at our cluster assignment predictors for the CGLM, while the mode assignments are taken ``as-is'' for the MGLM. \\

For the NDP, posterior medians, inter-quartile ranges (IQRs) and 95\% credible intervals are calculated for the intensity function parameters, $(\mu_{lk},\sigma^2_{lk})^*$, $\pi^*_k$ as well as the probability of co-cluster membership $\bm{P}$ as described in Section \ref{sec:mod_outcomes}. The $f(r)$ were constructed over a fine grid of equally spaced values in $\mathbb{R}$ representing the distances of a BEF from a school, combining the $K$ clusters and the $L$ within-cluster components at each distance. 
Since, for computational purposes, we transformed the school distances to FFRs from the interval $(0,R)$ to $\mathbb{R}$, when inferring upon the actual densities $f(r), r\in (0,R)$, we back-transform them onto the $(0,R)$ domain using the inverse probit function,  rescaling them by an empirically calculated proportionality constant. 

\par In the health outcome models, posterior median and 95\% credible intervals are calculated for regression coefficients $\bm{\beta}$, cluster effects $\xi_k, k=1,...,6$, and $h(\bm{P}_i)$, $i=1,\ldots,N$. Additionally, we calculate the effects of the quantity of FFRs as described in the following section.

\subsubsection{Quantity Effect}
\label{sec:prob_calc}

Given that schools with $n_i=0$ cannot be assigned a cluster for obvious reasons, the CGLM includes the  non-standard interaction terms between quantity $n_{i'}>0$ and cluster assignment.  As in any model with interactions, describing the ``main effect'' of an exposure requires careful attention. In this case, the ``main effect'' of the quantity of FFRs depends on the cluster to which schools with $n_{i'}>0$ were assigned. For each category indicator
 $Q_{i',m}$ with $m>0$, we marginalize over the cluster assignment to define the probability  of obesity given category $Q_{i',m}$; that is, the \emph{quantity effect}, holding $\bm{Z}_i=0$ as defined in Section \ref{sec:mod_outcomes} , is $$P(\text{Obesity} \;| \; Q_m,\text{Data}) = \sum_{k=1}^{K} w_k\text{inv-logit}(\zeta_m + \xi_k),$$ 
 where $w_k$ is the probability of a school being assigned to cluster $K$ in the $\mathbb{D}_{CGLM}$ dataset. Note that for $n_i=1$ , there is no corresponding effect, $\zeta_1$, as this is defined as the average cluster effect by the construction of the model in (\ref{eqn:outcomes_cglm}). 

Similarly, for the BKMR we calculate the FFR quantity effect on schoolchild obesity, now averaging over the $h(\bm{P}_i)$'s, using the following probability expression:
 $$P(\text{Obesity}\; | \; \zeta_m,\text{Data}) =  \frac{1}{\mid \mathbb{D}_{BKMR} \mid} \sum_{i=1}^{\mid\mathbb{D}_{BKMR}\mid} \text{inv-logit}(\zeta_m + \widehat{h(\mathbf{P}_i)}),$$ 
 where $\widehat{h(\mathbf{P}_i)}$ is the posterior mean of $h(\mathbf{P}_i)$.

\section{Results}

We now discuss results from both the clustering model, that aims to identify major patterns in the spatial distribution of FFRs in a 1 mile radius ($R=1$) around schools, and the models that relate the spatial distribution of FFRs to the odds of obesity in schoolchildren.

\subsection{Spatial Intensity Functions}

\par
\begin{figure}[H]
    \centering
    \includegraphics[width = .9 \textwidth ]{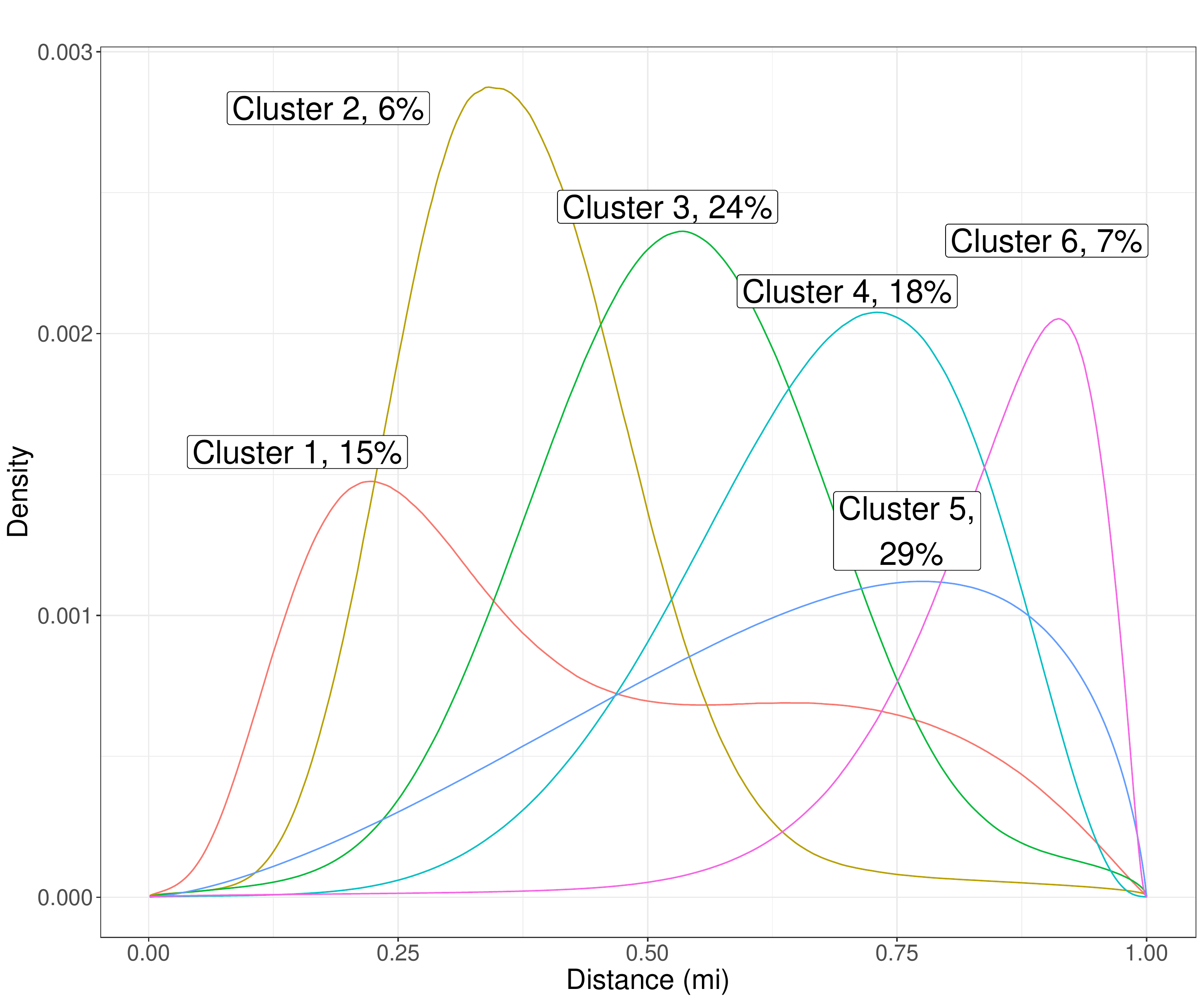}
    \caption{Estimate of cluster density functions $f^{*}_k(r)$, $k=1,\ldots,6$, with the estimated percent of schools within each cluster, $\pi^*_k$. The estimate here is taken to be the posterior median. The IQR for the percent of schools in each cluster are, for clusters 1 to 6, respectively: 3, 2, 4, 5, 5, and 2$\%$}
    \label{fig:est_clusters}
\end{figure}

 
The clustering model estimates six clusters with high probability, with the estimates of the cluster-assignment probabilities, $\pi^*_k$, beyond the first six effectively negligible when rounding to the hundredths place. The median density estimates, representing the likelihood of finding an FFR at a given distance from a school, are presented in Figure \ref{fig:est_clusters}, along with the proportion of schools in each cluster. As the figure shows, clusters are labeled according to their mode's proximity to the school, i.e. the cluster which estimates most FFRs to be located nearest to schools is labeled cluster 1, and so forth. A figure with the densities on the real line as estimated in the model along with 95\% credible intervals is shown in the Supplementary Material \citep{supp}.

\begin{figure}[H]
    \centering
    \includegraphics[width =  .80\textwidth ]{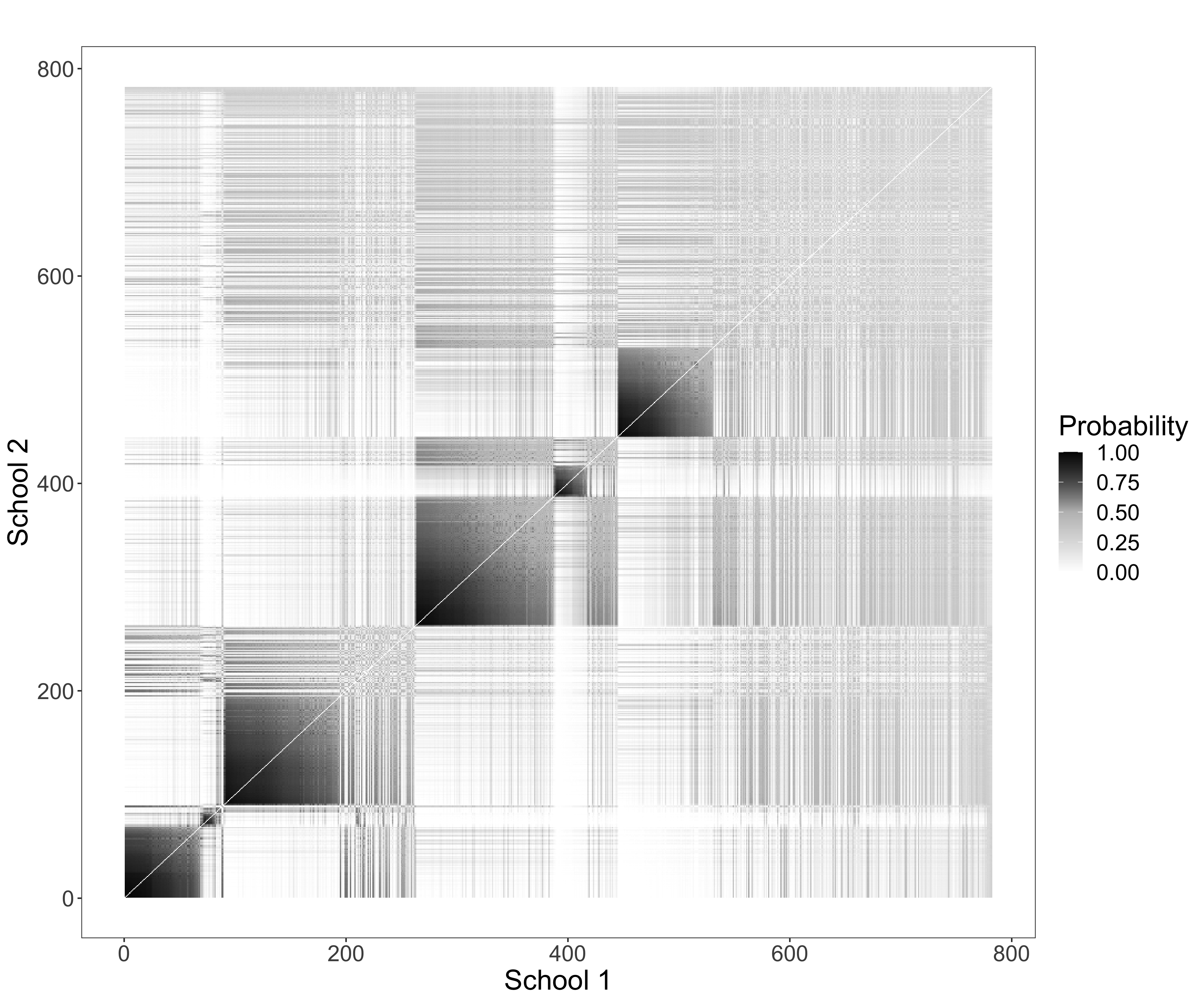}
    \caption{Heat map of co-clustering probabilities, that is, the probability that any two schools are assigned to the same cluster. The identity line may be interpreted as a school's probability of being clustered with itself. Although this probability is trivially equal to 1, for plotting purposes, in the figure this line is left equal to 0 to more clearly show the plot's line of symmetry.}
    \label{fig:pairwiseplot}
\end{figure}

In order to visualize how distinctly the clustering model assigns schools into the six different clusters, Figure \ref{fig:pairwiseplot} presents a heat map of the co-clustering probability matrix $\mathbf{P}$. Note that in the Figure, for the sake of visualization, school indices are arranged so that schools with similar co-clustering probabilities are next to one another, as implemented in the algorithm described in \cite{rodriguez2008nested}'s Supplementary Material. Examining the plot, we can clearly see the six clusters from left to right followed by the remaining schools which the model cannot cluster consistently.  

Table \ref{tab:cluster_table} presents summary statistics for the characteristics of the schools included in the six clusters identified, including a tabulation of the categorical variable describing the number of FFRs within a 1-mile of each school. In the table, we also include the characteristics of schools for high schools that have no FFRs within one mile of their location: we label this cluster as ``Cluster 0''.
There is a weak association between a school's census tract median household income and cluster membership. While Cluster 1's median census tract median income is \$55,200, Cluster 6's median census tract median income is \$67,400. However, this patterning does not include Cluster 0, which has a lower median census tract median income of \$53,900. A similar, though even weaker, pattern can be seen in the proportion of residents in the schools' census tracts with 16 or more years of education. Forty-four percent of schools in cluster 1 have  majority of white students populations, whereas 38 have predominately Latino students; for cluster 6, these percentages change to 58 and 16, respectively. Notably, all clusters contain schools across all urbanicity classification, and include schools with a varying number of FFRs. In other words, the mode cluster is not driven by FFR quantity or broader context (e.g., urbanicity) of the schools. 

\begin{landscape}
\begin{figure}[H]
    \centering
    \includegraphics[width =  1.15\textheight]{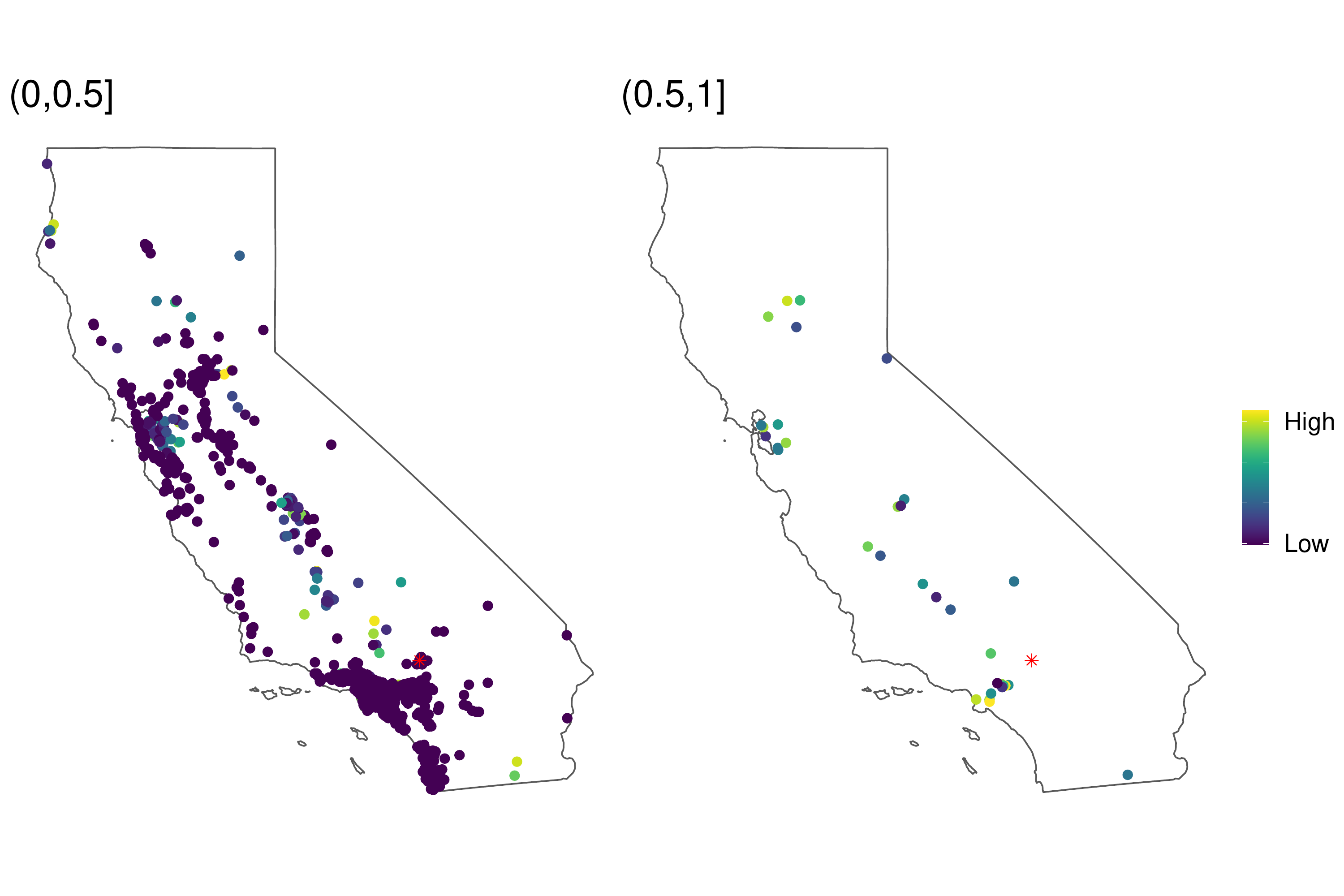}
    \caption{Map of the probability of co-clustering with the school denoted by a star. Probabilities are color-coded with lighter colors indicating larger probabilities within each of 2 probability intervals considered, (0,0.5] and (0.5,1]. }
    \label{fig:probmap}
\end{figure}
\end{landscape}

To assess whether the six identified clusters were geographically concentrated in one or more sub regions of California, and to investigate whether schools tended to co-cluster with nearby schools,  we produced spatial plots of the co-clustering probabilities for a given school. Figure \ref{fig:probmap} presents this plot for a school located in Southern California, identified in the map by a star symbol. As the figure shows, the schools that are more likely to be co-clustered with the selected school are not necessarily located nearby. Rather, as the first panel of the figure shows, most of the schools nearby the chosen school tend to have a probability smaller than 0.5 of being assigned to the same cluster.
\begin{landscape}
\begin{table}[H]
\centering 
\small
\begin{tabular}{llcccccccc}
\toprule
& & \multicolumn{7}{c}{Mode Cluster} & \multicolumn{1}{c}{} \\ 
 &  & 0 & 1 & 2 & 3 & 4 & 5 & 6 & \multicolumn{1}{c}{Total} \\ 
 & & (n=426) & (n=103) & (n=28) & (n=231) & (n=105) & (n=252) & (n=31) & (n=1176) \\
\midrule
FFR Quantity within 1 mile & [1,4] &  $\phantom{0}0$ & $42$ & $39$ & $48$ & $34$ & $47$ & $55$ & $29$ \\
 & $\geq$5 &  $\phantom{0}0$ & $58$ & $61$ & $52$ & $66$ & $53$ & $45$ & $35$ \\
 & Zero &  $100$ & $\phantom{0}0$ & $\phantom{0}0$ & $\phantom{0}0$ & $\phantom{0}0$ & $\phantom{0}0$ & $\phantom{0}0$ & $36$ \\
Urbanicity & Rural &  $39$ & $10$ & $14$ & $13$ & $10$ & $\phantom{0}6$ & $19$ & $21$ \\
 & Sub-Urban &  $34$ & $40$ & $39$ & $44$ & $44$ & $46$ & $42$ & $40$ \\
 & Urban &  $27$ & $50$ & $46$ & $43$ & $47$ & $48$ & $39$ & $39$ \\
Majority Race/ethnicity  & African American &  $\phantom{0}2$ & $\phantom{0}2$ & $\phantom{0}4$ & $\phantom{0}1$ & $\phantom{0}0$ & $\phantom{0}0$ & $\phantom{0}3$ & $\phantom{0}1$ \\
among enrolled students & Asian &  $\phantom{0}2$ & $\phantom{0}6$ & $\phantom{0}4$ & $\phantom{0}4$ & $\phantom{0}4$ & $\phantom{0}5$ & $\phantom{0}3$ & $\phantom{0}4$ \\
 & Hispanic &  $27$ & $38$ & $21$ & $27$ & $29$ & $29$ & $16$ & $28$ \\
 & No Majority &  $\phantom{0}9$ & $11$ & $14$ & $13$ & $18$ & $13$ & $19$ & $12$ \\
 & White &  $59$ & $44$ & $57$ & $55$ & $50$ & $52$ & $58$ & $55$ \\
Median Income (1,000 USD) & Median   & $53.9$ & $55.2$ & $55.7$ & $61.0$ & $69.7$ & $61.1$ & $67.4$ & $58.6$ \\
 &(Q1-Q3)  &  (41.8-76) &  (44-77.2) &  (45.4-75.7) &  (44.1-83.2) &  (48.9-90.6) &  (47-77) &  (43.4-86.5) &  (44.2-79.3) \\
& IQR  & $34.1$ & $33.3$ & $30.2$ & $39.1$ & $41.7$ & $30.1$ & $43.0$ & $35.1$ \\
Proportion of adults & Median & $24.9$ & $25.0$ & $25.4$ & $25.4$ & $25.6$ & $25.3$ & $25.5$ & $25.2$ \\
with $\geq$ 16 years of education & (Q1-Q3) &  (24.1-26.2) &  (24.2-26.9) &  (24.1-27.5) &  (24.4-27) &  (24.3-27.2) &  (24.1-26.7) &  (24.5-26.9) &  (24.2-26.6) \\
  & IQR  & $\phantom{0}2.1$ & $\phantom{0}2.7$ & $\phantom{0}3.4$ & $\phantom{0}2.6$ & $\phantom{0}2.9$ & $\phantom{0}2.6$ & $\phantom{0}2.5$ & $\phantom{0}2.4$ \\
\bottomrule
\end{tabular}
\caption{Descriptive statistics of school characteristics by mode cluster assignment. Summary statistics - percents, median and inter-quartile range (IQR) for categorical and continuous school-level or census-tract level covariates for each cluster. In the table, the column designated as "Cluster 0" reports summary statistics for those high schools without any fast food restaurants within one mile of their location.``Median Income'' and ``Proportion of residents'' refer to characteristics of the population living in the census tract in which schools are located.}
\label{tab:cluster_table}
\end{table}

\end{landscape}

\subsection{Health Outcomes Models}

In discussing the results of our proposed health outcomes models, we start by providing a description of the schools whose data are used. The proportion of students that are obese is similar in both the consensus and full datasets (Table \ref{tab:conensus}), which is encouraging since schools are not excluded on the basis of the outcome in the consensus dataset. However, schools used to fit the consensus model are less likely to have few FFRs around them - only 21\% have between 1 and 4 FFRs compared to 45\% in the full dataset. 

Turning to outcome model results, we'll discuss both second stage approaches on both the consensus and full datasets, starting with the consensus dataset. However, since the BKMR results mirror those of the CGLM, we'll focus on how these second-stage models reinforce one another rather than describing each individually.

\par
\begin{figure}[H]
    \centering
    \includegraphics[width = .8\textwidth]{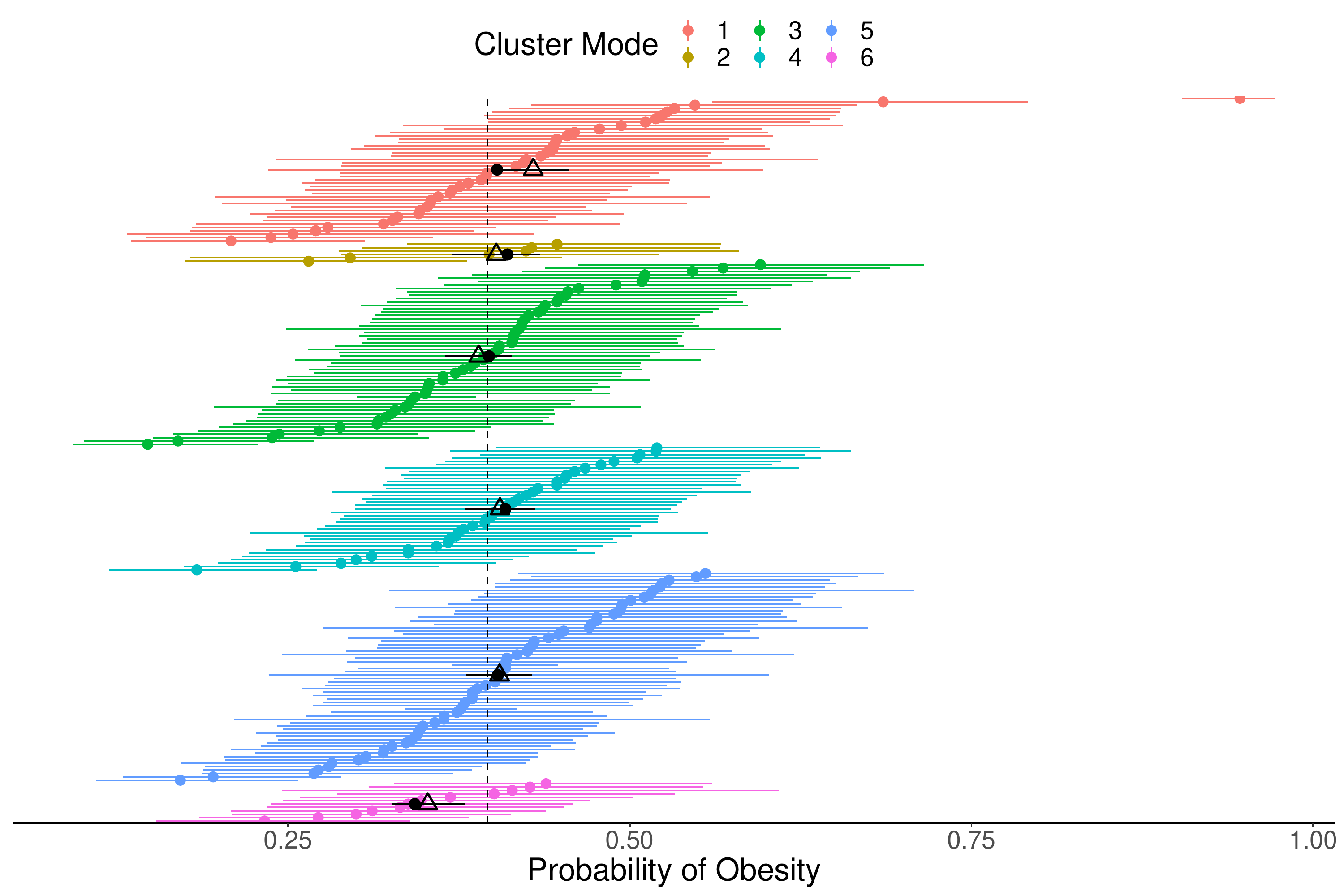}
    \caption{Probability of obesity in relation to fast food restaurant (FFR) proximity. Estimates from the
  Bayesian Kernel Machine Regression (BKMR) are shown for each school (dot), along with 95 \% credible intervals (line), and are colored according to the cluster mode assignment. Dark dot represents the overall median probability of obesity for children attending schools in the given cluster.   Triangles (and horizontal black line) denote the median posterior probability of obesity for children attending schools in each cluster estimated from the consensus GLM (CGLM) along with the 95\% credible interval interval. The reference dotted vertical line is the posterior mean probability of obesity at a majority White sub-urban high school with at least one FFR within a mile of the school's location. BKMR and CGLM results are estimated using the consensus data set.}
    \label{fig:second_stage_consensus}
\end{figure}

 As shown in Figure \ref{fig:second_stage_consensus}, we observe a monotonic decrease in the probability of obesity as a function of the proximity of FFRs to the school, after adjusting for 1 mile radius quantity of FFRs. Specifically, according to the CGLM, children attending schools consistently assigned to cluster 6 have a 35\% (95\% CI: 33\%,38\%) probability of being overweight or obese, while, for other clusters, the lower bound estimate of the probability of obesity ranges from 37\% to 40\%.    
These results are consistent with the substantive expectation that students who are exposed to FFRs in the immediate environment around schools are more likely to be obese than they would be otherwise. As  Figure \ref{fig:est_clusters} shows, the density of FFRs for schools in cluster 6 is greatest after 3 quarters of a mile, in explicit contrast to the other clusters which tend to have greater density of FFRs closer to the school. This finding supports prior work suggesting that zoning laws that restrict the placement of fast food restaurants could serve as possible population-level strategies to reduce child obesity \cite{austin2005clustering}.  

Figure \ref{fig:second_stage_consensus} overlays the results of the CGLM and the BKMR models, demonstrating their agreement. The figure also shows that the BKMR provides additional information regarding the probability of obesity for children in each school. Beyond potential policy implications of the average obesity risk for children across the school food environment clusters, the school-level estimates can be used to prioritize individual schools for additional interventions.

\par
\begin{figure}[H]
    \centering
    \includegraphics[width = \textwidth]{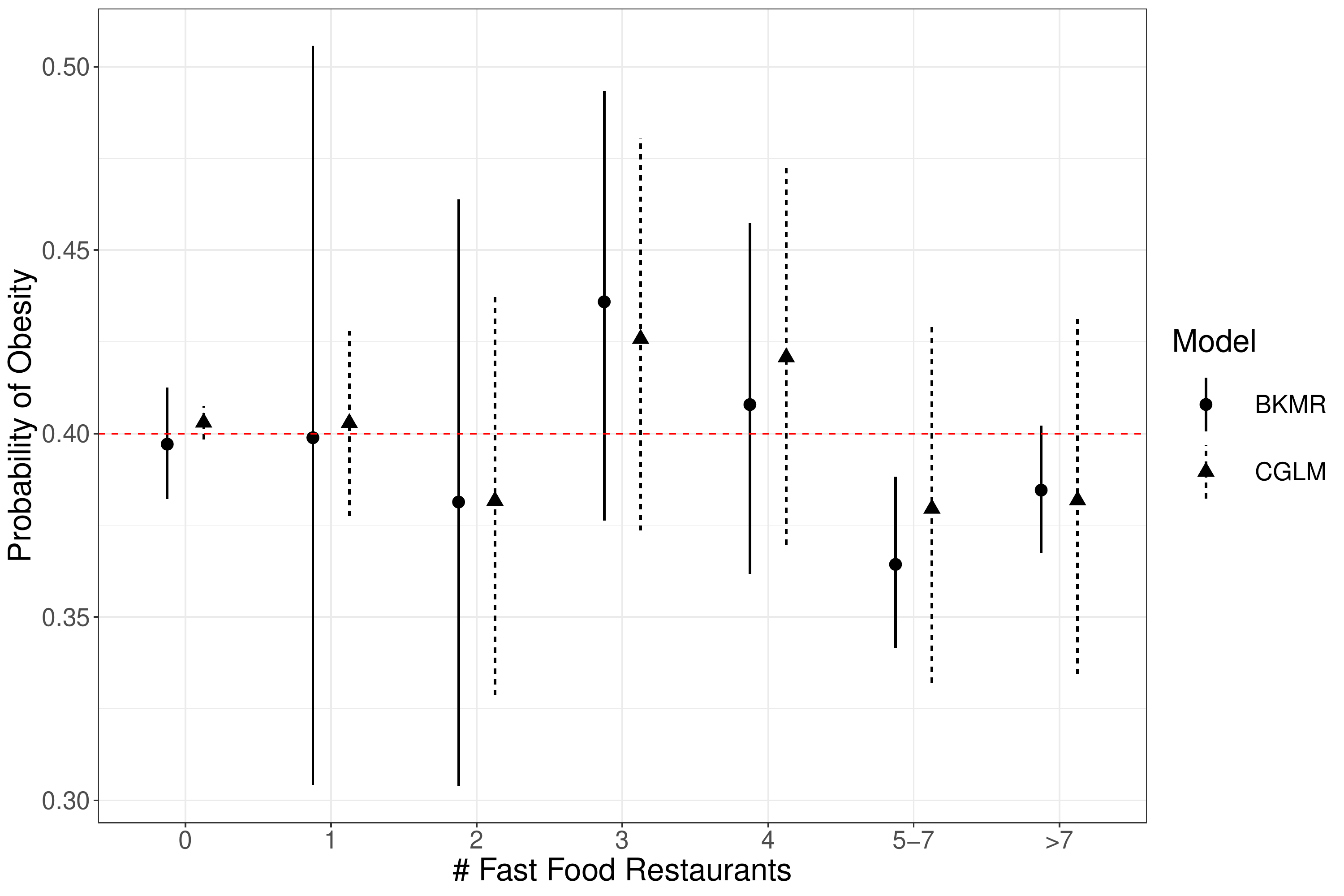}
    \caption{Posterior probability of obesity  and 95\% credible intervals according to the number of FFRs surrounding a school, adjusted for the effect of proximity of FFRs. Results refer to analysis performed on the  consensus dataset, as estimated by the two models.}
    \label{fig:second_stage_quantity}
\end{figure}

  Figure \ref{fig:second_stage_quantity} shows the estimated probability of obesity as a function of the number of FFRs within a 1-mile radius of the schools, calculated as described in Section \ref{sec:prob_calc}.  As the figure indicates, there is a general agreement between the CGLM and BKMR models with respect to the negligible effect of the number of FFRs on obesity after adjusting for the FFRs' spatial distribution and other covariates. The only estimate that stands out from these analyses is the BKMR's estimate of lower obesity for children in schools with 5-7 FFRs nearby, as compared to zero FFRs - a counter intuitive result. However, it is possible that the greater number of FFRs implies greater variety of food choices, including healthier options. The data set in this analysis does not contain information on the specific types of FFRs, beyond the number and location, thus not allowing us to examine this possibility.

As with the consensus data set, the results from fitting a GLM (using only the median cluster assignment) and the BKMR in the full data set are in agreement with each other, as shown in Section 1 of the Supplementary Material \citep{supp}. However, the results from the analysis on the full data set instead identify Cluster 2 as having the lowest probability of obesity, at 37\% (95\% CI: 36\%,38\%), with the probability for all other clusters near or above 40\%. Differences in the association between the spatial distribution of FFRs near schools on child obesity, comparing the full and consensus data sets are likely due to the fact that the full dataset contains schools with more uncertain cluster assignments, and thus potentially more prone to miss-classification errors and thus bias in the associations. The quantity effects are similar in the consensus and full data set, and again agree between methods (see Figure 3 in Section 1 of the Supplementary Material).

Finally, comparison of these models to more traditional competitor models by WAIC are shown in Supplementary Section 4, Table 1 \citep{supp}. Both the BKMR and CGLM perform better by this metric than their more traditional counterparts regardless of what dataset they have been fit to.

\begin{table}[H]
\centering
\begin{tabular}{llccc}
\hline
 &  &   In Consensus & Not in Consensus & \multicolumn{1}{c}{All} \\ 
\hline
Proportion Obese & Median (Q1-Q3)  & 40.9 (33.3-47.4) & 41.3 (34.1-48.2) & 41.3 (33.9-48) \\
 & IQR  & $14$ & $14.1$ & $14.2$ \\
\hline 
FFR Quantity within 1 mile & [1,4] &  $21$ & $55$ & $45$ \\
 & $\geq$5 &  $79$ & $45$ & $55$ \\
Urbanicity & Rural &  $\phantom{0}8$ & $11$ & $10$ \\
 & Sub-Urban &  $35$ & $47$ & $44$ \\
 & Urban &  $57$ & $42$ & $46$ \\
Majority Race & African American &  $\phantom{0}1$ & $\phantom{0}1$ & $\phantom{0}1$ \\
among enrolled students & Asian &  $\phantom{0}5$ & $\phantom{0}4$ & $\phantom{0}4$ \\
 & Hispanic &  $30$ & $28$ & $29$ \\
 & No Majority &  $16$ & $13$ & $14$ \\
 & White &  $49$ & $53$ & $52$ \\
Median Income (1,000 USD) & Median (Q1-Q3)  & 60.4 (43.2-78.4) & 61.4 (46.2-82.9) & 61.2 (45.3-81.5) \\
 & IQR  & $35.2$ & $36.7$ & $36.2$ \\
Proportion of residents  & Median (Q1-Q3)  & 25.3 (24.3-26.7) & 25.4 (24.2-27) & 25.4 (24.2-26.9) \\
 with $\geq$ 16 years of education & IQR  & $\phantom{0}2.5$ & $\phantom{0}2.7$ & $\phantom{0}2.7$ \\
\hline 
\end{tabular}

\caption{Descriptive Statistics for Schools Analyzed using the Consensus GLM vs. not. IQR = Inter-quartile range; FFR = Fast Food Restaurant. All numerical values for categorical rows are the column percentage within the left row heading. ``Median Income'' and ``Proportion of residents...'' are characteristics of the population living in the census tract where schools are located. }
\label{tab:conensus}
\end{table}

\section{Discussion}
In this work we have presented a two-stage modeling strategy that aims to provide epidemiological and social science investigators with a tool that permits them to both identify links between exposure to specific features of the built environment and health outcomes as well as identify those subjects at greatest risk of negative health outcomes. In the first stage, our goal is to identify major patterns in the spatial distribution of FFRs around a school and group schools together based on their surrounding food environment. The second stage links the spatial distribution of FFRs around schools to the likelihood that children in the schools are obese. This work can be easily adapted to answer questions involving the association between other point-referenced amenities in the built environment and health outcomes, for example, availability of parks and measures of physical activity \citep{evenson2016park}, depression \citep{bojorquez2018urban}, or availability of social engagement destinations and cognition, among others \citep{besser2018neighborhood}.

\par Our work breaks with previous approaches to quantify exposures to the built environment in several ways.  First, we use a point pattern approach to model exposure to BEFs, which is not typically done in this type of literature.  Second, our clustering model differs from other parametric clustering models used in built environment research in that no constraints are imposed on the number of clusters. Third, in addition to selecting the number of clusters in a data-adaptive fashion, our model estimates the cluster specific densities non-parametrically, thanks to the use of the NDP prior on the density function representing the distances between a given school and nearby FFRs. This modeling strategy allowed us to identify clusters of high schools in California that have a high number of FFRs in their immediate environment relative to their peers, and those that have FFRs farther way. 

\par Our work also proposes approaches to incorporate output from a clustering method into a second stage regression model; namely, by using either a decision-theoretic framework to control uncertainty in cluster assignment, or using the posterior co-clustering probability matrix to borrow information across observations without needing to reduce the exposure information to discrete exposure groups. The latter approach extends the uses of kernel machine regression to applications in the built environment, beyond those used to examine health effects of chemical contaminants \citep{bobb2015bayesian}. Through our proposed second-stage health outcome models, we incorporated information on the spatial distribution of point-pattern amenities into a model for child obesity. These approaches identified that, independent of the quantity of FFRs within a mile of California high schools, children attending schools with FFRs further away had lower obesity risk. Though the differences in obesity risk between food environment clusters were relatively small, it is well understood that ubiquitous exposures, even if they have small individual-level effects, can result in large population health impacts \citep{rose2001sick}.

\par The second stage models have different advantages, disadvantages and ways of incorporating BEF exposure information obtained from the first stage. Specifically, the CGLM has the potential to suffer from selection bias  by using only the subjects with highest certainty in their class assignment.  In our analysis, although the schools with higher uncertainty tended to have fewer outlets nearby, the excluded schools did not differ in terms of the outcome, thus minimizing selection bias concern upon conditioning by the number of FFRs in the second stage.  Users of the CGLM should keep in mind this potential for selection based on the outcome, in which case inverse probability of selection into the second stage could be used. The BKMR can use all subjects, handles cluster assignment uncertainty by using the co-clustering probability matrix, and can provide more granular information about health outcome risk for each subject in a school through the posterior estimates of $h(\mathbf{P}_i)$. However, visualizing/interpreting the BKMR's rich set of output could be challenging.  In our case, our goal was to compare the results of the analyses between the two methods  and thus we used the mode cluster label to visualize the BKMR results. Other visualizations of the results may include displays of plots of the $\widehat{h(\bm{P}_i)}$ as a function of the L$^2$ norm of the co-clustering probabilities, $\bm{P}_i$ and $\bm{P}_j$ for a reference school $j$.

Pursuing methods that more comprehensively estimate or propagate the uncertainty associated with cluster assignment from a DP clustering approach through the health outcome analysis may be desirable, as neither the BKMR or CGLM fully do so. One possible solution would be to develop a method that allows for joint estimation of both the cluster specific densities as well as the cluster-associated outcome risk.  Our current method is unable to easily embrace such a joint modeling approach due to to both label switching and the varying number of assigned clusters across the MCMC iterations, yielding identifiability problems for the health outcomes models \citep{gelman2013bayesian}. This makes the goal of joint estimation more difficult and a promising subject of future work. One possible solution would be to incorporate the health outcome at the level at which the cluster is constructed. Adapting the Logistic Stick Breaking Process \citep{ren2011logistic} for example, could facilitate this goal.  Nevertheless, we emphasize that, while the two-stage approach proposed here does not propagate uncertainty in cluster assignment in a standard fashion, this strategy still offers a number of benefits. For example, defining the exposure clusters independent of the outcome ensures a greater level of interpretability and conforms to substantive understanding of such clusters \citep{nylund2019prediction}. Furthermore, it offers a greater level of applicability for the clusters, as estimating them separately from the outcome means they can be used for more than one health outcome analysis.
 
  \par 
 Finally, we acknowledge two recent theoretical results pertaining to the NDP and DP, respectively. In the first, \citet{camerlenghi2019latent} showed that the NDP estimates of two or more distributions can collapse or degenerate to a single estimated distribution in the case when there are ties in the observations across subjects (e.g. two schools that have the exact same school-FFR distance) or when the true underlying distributions for different clusters share atoms.  In our case, this can lead to collapsing or merging of intensity function estimates and thus potentially a lower number of school clusters identified. In the case of ties, we believe this should not be of substantial concern in the application that we are proposing our methodology for, as ties rarely occur in the calculation of distances at sufficient precision (none occurred in our motivating dataset) and, if ties do occur, a small amount of normal random error can be added to the distances to quickly resolve this issue without significantly affecting inference. The question of whether a latent mixture component is shared between two normalized intensity functions is more difficult to answer. However, we believe that the variability in these kinds of data, as illustrated by Figure \ref{fig:composite_distance}, provides evidence that this may be of less concern here. If a dataset exhibits less variability than the one examined here, then greater caution may be warranted. 
 
 \par 
 The second theoretical result of note showed that the DP cannot consistently estimate the number of clusters when the concentration parameter is fixed \citep{miller2013simple,miller2014inconsistency}. The same authors suppose a similar result holds in the more general case of a random concentration parameter, though there is as yet no proof \citep{miller2018mixture}.  The challenges this unknown feature presents can be accommodated by the two different outcome models we presented. If one is unconcerned about the potential lack of consistency and willing to rely on the concentration parameter to correctly inform the appropriate number of clusters, then the CGLM will offer a standard interpretation that relies on the number of clusters being a consistent estimate of the truth. In contrast, should there be concern that the NDP cannot consistently estimate the correct number of clusters then the BKMR offers a better approach -- as it does not rely on the concept of their being some definite number of clusters, but rather uses the matrix of co-clustering probabilities to provide information about differing levels of exposure. 

\par
Further extensions of the work presented should incorporate the spatial distribution of more than one BEF amenity. The built environment consists of many amenities beyond FFRs that could be co-located with FFRs, or have different spatial distributions leading to different 'mixtures' of amenities. Extending our model to higher dimensions could allow investigators to characterize joint exposure to multiple amenities and identify their corresponding relationship with health outcomes. In addition, model extensions that incorporate the spatial proximity of subjects, in our case schools, when forming clusters would be of interest. In our analyses, we selected schools that were far apart from one another to satisfy independence assumptions.  The work presented here represents a first step in building relevant descriptions of the built environment in which humans live and informing decisions as to how new built environments may be constructed in the future.

%%%%%%%%%%%%%%%%%%%%%%%%%%%%%%%%%%%%%%%%%%%%%%
%% Single Appendix:                         %%
%%%%%%%%%%%%%%%%%%%%%%%%%%%%%%%%%%%%%%%%%%%%%%
%\begin{appendix}
%\section*{???}%% if no title is needed, leave empty \section*{}.
%\end{appendix}
%%%%%%%%%%%%%%%%%%%%%%%%%%%%%%%%%%%%%%%%%%%%%%
%% Multiple Appendixes:                     %%
%%%%%%%%%%%%%%%%%%%%%%%%%%%%%%%%%%%%%%%%%%%%%%
%\begin{appendix}
%\section{???}
%
%\section{???}
%
%\end{appendix}

%%%%%%%%%%%%%%%%%%%%%%%%%%%%%%%%%%%%%%%%%%%%%%
%% Support information (funding), if any,   %%
%% should be provided in the                %%
%% Acknowledgements section.                %%
%%%%%%%%%%%%%%%%%%%%%%%%%%%%%%%%%%%%%%%%%%%%%%
\section*{Acknowledgements}

This research was partially supported by NIH grants R01-HL131610 (PI: S\'anchez) and R01-HL136718 (MPIs: Sanchez-Vaznaugh and S\'anchez).

% 
% The second author was supported in part by ...
 
%%%%%%%%%%%%%%%%%%%%%%%%%%%%%%%%%%%%%%%%%%%%%%
%% Supplementary Material, if any, should   %%
%% be provided in {supplement} environment  %%
%% with title inside \textbf{} and short    %%
%% description below.                       %%
%%%%%%%%%%%%%%%%%%%%%%%%%%%%%%%%%%%%%%%%%%%%%%
\begin{supplement}
\stitle{Supplementary Information}
\slink[doi]{COMPLETED BY THE TYPESETTER}
\sdatatype{.pdf}
\sdescription{We provide additional supporting plots and tables that show 1. The Estimated Intensity Functions on the real line. 2. Mode GLM and BKMR results fit to full dataset 3. Posterior mode vs consensus glm probability of obesity estimates. 4. Table of WAIC values comparing traditional methods to our proposed methods for both full and consensus datasets. 5. A table featuring further descriptive statistics of relevant school-level covariates. }
\end{supplement}

%%  imsart-nameyear.bst  will be used to                   %%
%%  MR numbers will be added by VTeX.                      %%
%%                                                         %%
%%  Use \cite{...} to cite references in text.             %%
%%                                                         %%
%%%%%%%%%%%%%%%%%%%%%%%%%%%%%%%%%%%%%%%%%%%%%%%%%%%%%%%%%%%%%

%% if your bibliography is in bibtex format, uncomment commands:

% Style BST file
\bibliographystyle{imsart-nameyear} 
% Bibliography file (usually '*.bib')
\bibliography{references}

\end{document}

\maketitle
% section 1
\section{Real-Line Intensity Function Estimates}
\begin{figure}[H]
    \centering
    \includegraphics[width = .85 \textwidth]{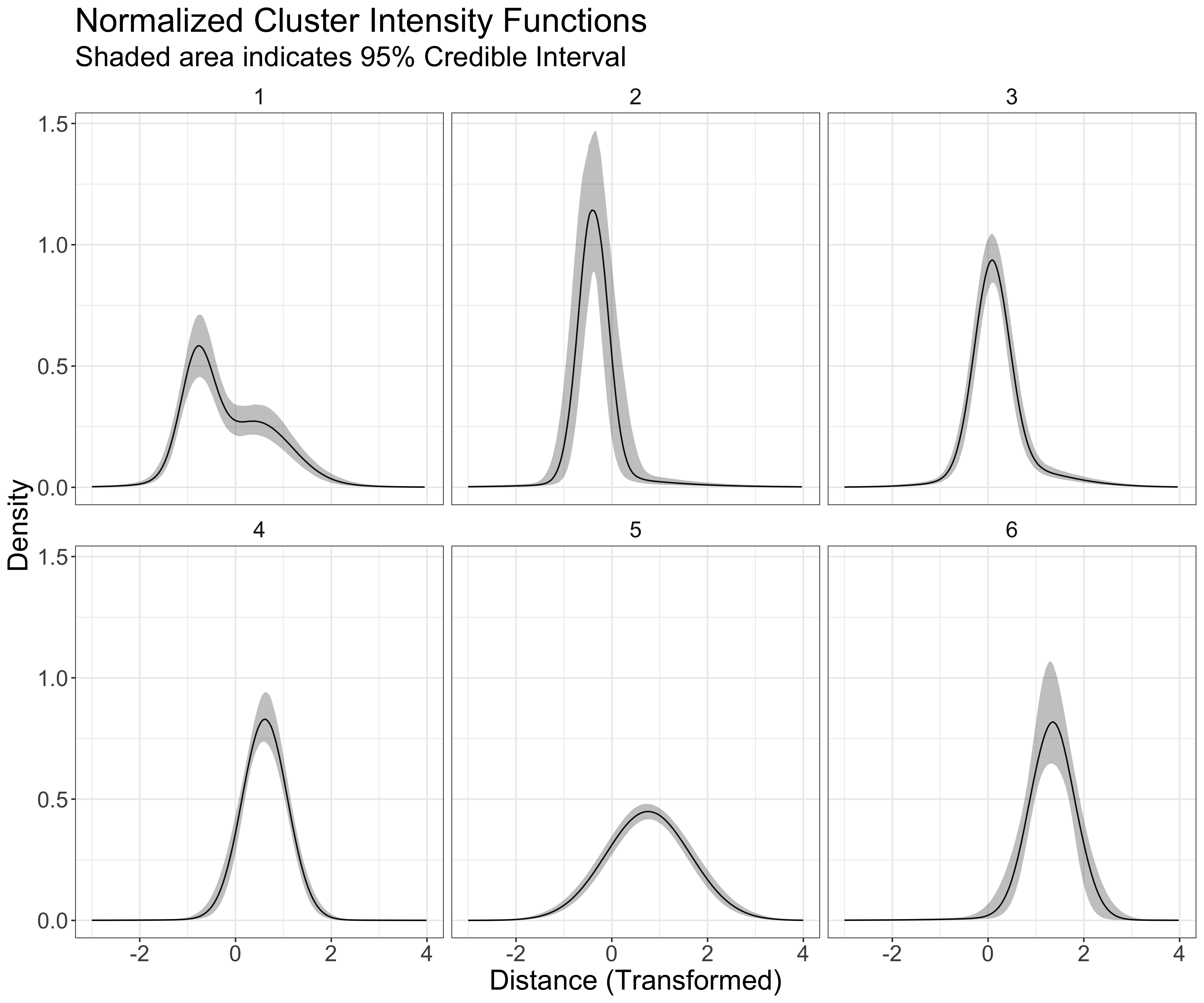}
    \caption{Median and 95\% Credible Interval Estimates for cluster normalized intensity functions on transformed $\mathbb{R}$ scale.}
    \label{fig:raw_cluster_intensities}
\end{figure}
% section 2
\section{BKMR and Mode GLM Results on the Full Dataset}
\begin{figure}[H]
    \centering
    \includegraphics[width = \textwidth]{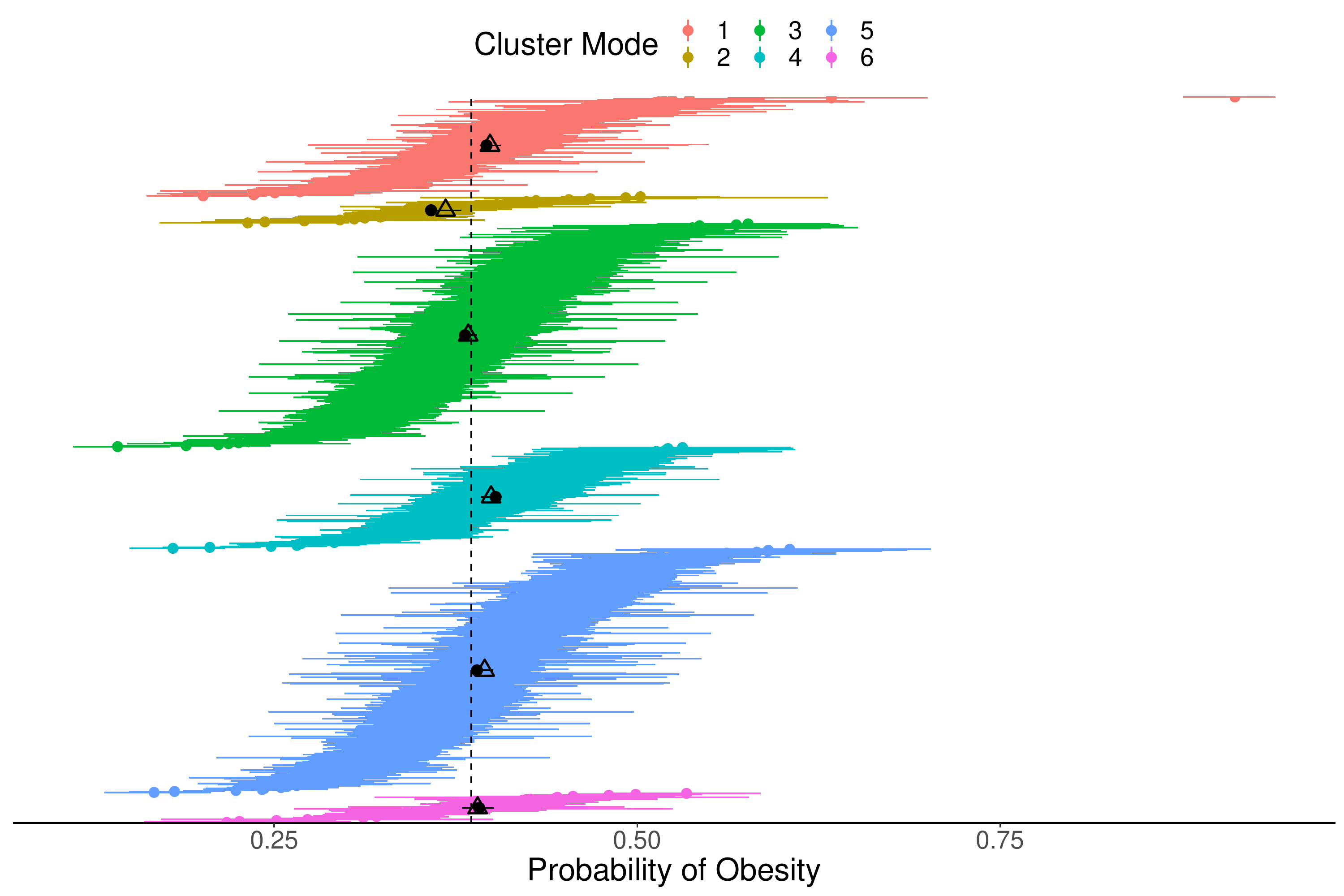}
    \caption{Health Outcome Fast Food Restaurant (FFR) Spatial Proximity Effects. Bayesian Kernel Machine Regression (BKMR) random school intercepts 95 \% credible interval are plotted as lines with colored cluster median dots. Mode GLM (MGLM) effects' 95\% credible intervals for each cluster are plotted with triangles denoting the median estimate. The reference dotted line is the posterior mean probability of obesity for children in suburban high schools with a majority of white students, with at least one FFR within a mile of the school's location. BKMR and MGLM results are estimated from a datasets of 1176 schools.}
    \label{fig:second_stage_full}
\end{figure}

\begin{figure}[H]
    \centering
    \includegraphics[width = \textwidth]{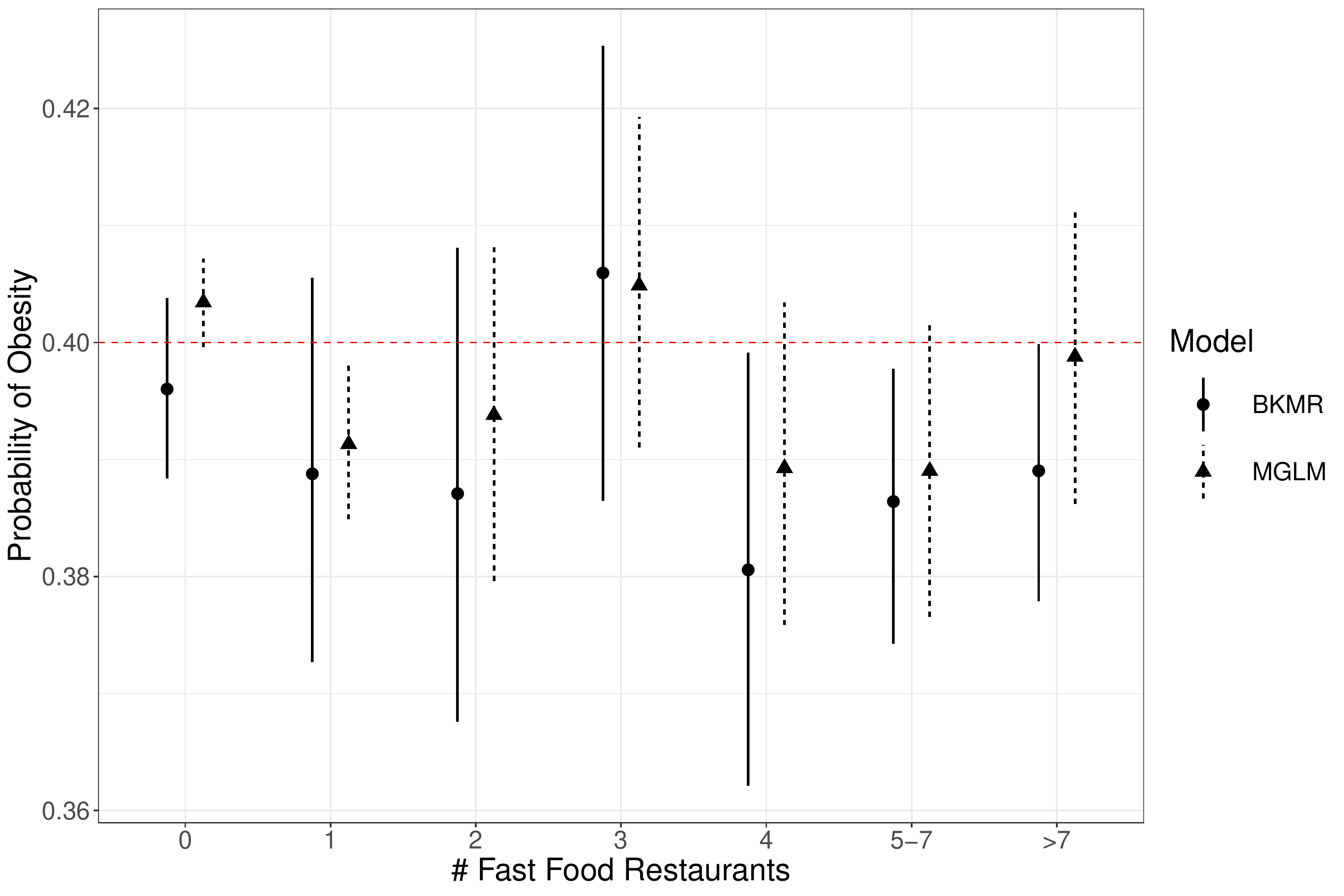}
    \caption{Health Outcome fast food restaurant Quantity Effect from full dataset. Point ranges represent median and 95\% credible intervals}
    \label{fig:second_stage_quantity_full}
\end{figure}
% section 2
\section{Mode vs. Consensus GLM}

\begin{figure}[H]
    \centering
    \includegraphics[width = .85 \textwidth]{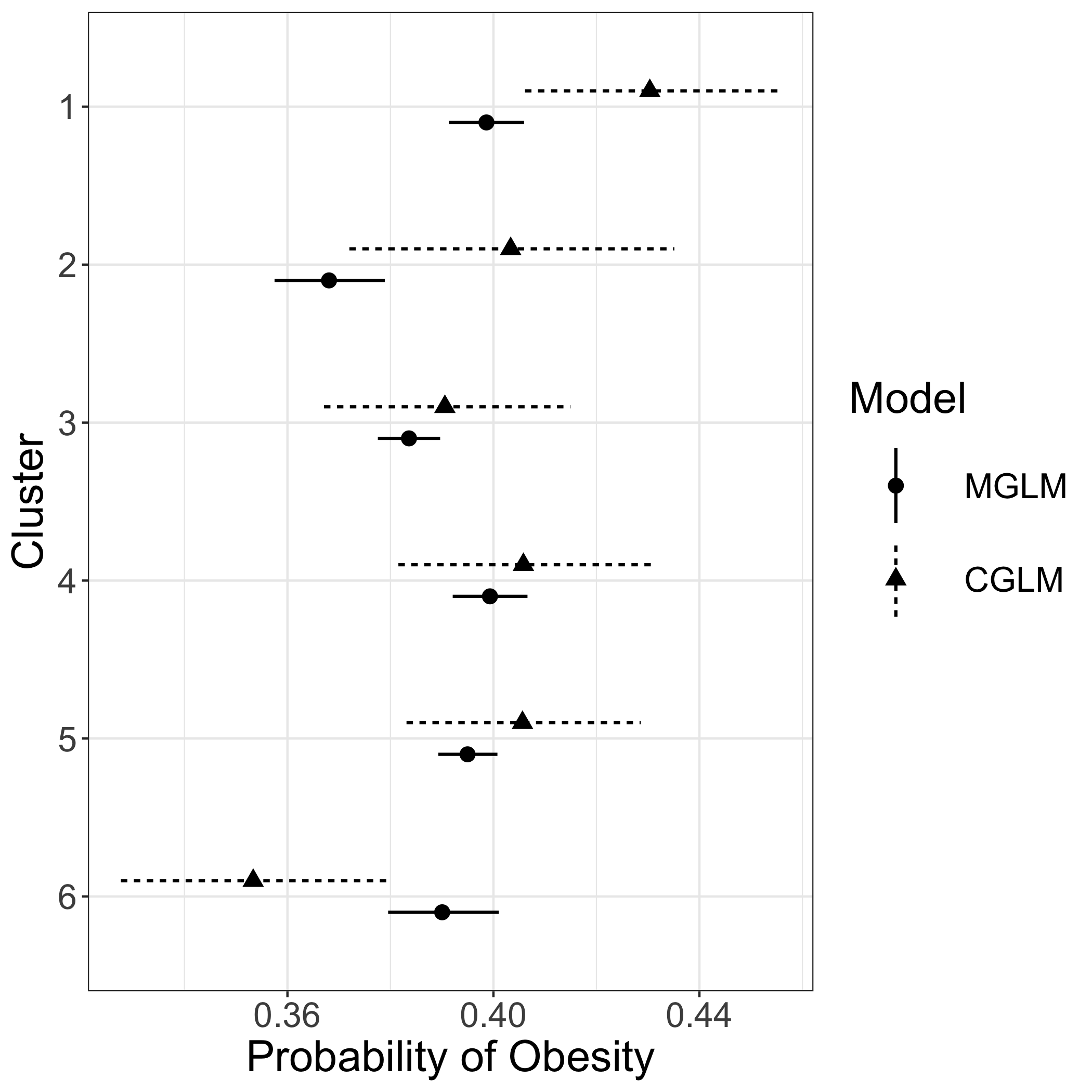}
    \caption{Posterior Mode (MGLM) and Consensus GLM (CGLM) analyses. Results show the school's proportion of obese students within each cluster configuration.}
    \label{fig:mode_glm_comp}
\end{figure}

% section 3
\section{ Preliminary Analysis: Model Comparison}
\begin{table}[H]
  \centering
    \begin{tabular}{lcc}
  \hline
Models & Full & Consensus \\ 
  \hline
T.1 & 30,566.95 & 12,217.12 \\ 
  T.2 & 33,972.94 & 17,040.08 \\ 
  T.3 & 26,096.42 & 12,169.42 \\ 
  BKMR & 11,126.52 & 6,922.43 \\ 
  CGLM & - & 9,883.43 \\ 
  MGLM & 17,612.0 & - \\
   \hline
\end{tabular}
    \caption{Widely Applicable Information Criterion (WAIC) for Traditional (T) models 1-3, Bayesian Kernel Machine Regression (BKMR) and Consensus GLM (CGLM) for both Consensus and Full datasets corresponding to ``In Consensus'' and ``All'' columns from Table 3, respectively. Each model contains the same adjusting covariates and different measures of FFR exposure in a logistic regression modeling 9th grader obesity. T. 1 includes the \# of FFR within 1 mile of the school. T. 2 includes the distance to the closest FFR and T. 3 includes both the previous measures. CGLM, MGLM and BKMR are as denoted in the text. }
    \label{tab:prelim_comparison}
    \end{table}
% section 4

\section{Descriptive Statistics}
\begin{table}[H]
    \centering
    \begin{tabular}{llccc}
\hline
 &  & $\geq 1$ FF & no FFRs nearby & \multicolumn{1}{c}{All schools} \\ 
\hline
\% Receiving Free   & Median (Q1-Q3)  & 48.9 (26.1-68) & 48.9 (27.6-71.9) & 48.9 (26.9-68.8) \\
 or Reduced Price Meals & IQR  & $41.9$ & $44.2$ & $42$ \\
\hline 
Traditional High School & & $78.13$ & $78.25$ & $78.18$ \\
Charter High School & & $21.87$ & $21.75$ & $21.82$ \\
\hline 
\end{tabular}
    \caption{Supplemental School Covariate Information. The lower half of the table contains the column percentage of schools that are either Charter or traditional.}
    \label{tab:supp_descriptive}
\end{table}